\documentclass[12pt,preprint]{aastex}
\usepackage{epsfig}

\newcommand{\eqb}{\begin{eqnarray}}
\newcommand{\eqe}{\end{eqnarray}}
\newcommand{\melec}{m_{\rm e}}
\newcommand{\mprot}{m_{\rm p}}
\newcommand{\gammap}{\gamma_{\rm p}}

\newcommand{\gammae}{\gamma_{\rm e}}

\newcommand{\nelec}{n_{\rm e}}
\newcommand{\nphotre}{n_\gamma^{\rm R}}
\newcommand{\nelsc}{n_{\rm e}^{\rm sc}}
\newcommand{\nprot}{n_{\rm p}}

\newcommand{\nphot}{n_\gamma}
\newcommand{\nphotref}{n_\gamma^R}
\newcommand{\bmin}{B_{\rm min}}

\newcommand{\Rbrw}{R_{\rm RBW}}
\newcommand{\sigmaT}{\sigma_{\rm T}}

\newcommand{\lphot}{\ell_{\gamma}}
\newcommand{\lphotref}{\ell_{\gamma}^R}
\newcommand{\bcomp}{\ell_{\rm b}}

\newcommand{\bcrit}{B_{\rm cr}}   
\newcommand{\protlopi}{L_{\rm p}^{\pi}}
\newcommand{\elinpi}{Q_{\rm e}^{\pi}}
\newcommand{\phinpi}{Q_{\gamma}^{\rm pi}}
\newcommand{\qbhinj}{Q_{\rm e}^{\rm BH}}
\newcommand{\qbhinjref}{Q_{\rm e}^{\rm BH,R}}
\newcommand{\protloss}{L_{\rm p}^{\rm BH}}
\newcommand{\protlossref}{L_{\rm p}^{\rm BH,R}}
\newcommand{\qsyninj}{Q_{\gamma}^{\rm syn}}
\newcommand{\lsynlos}{L_{\rm e}^{\rm syn}}
\newcommand{\qicsinj}{Q_{\gamma}^{\rm ics}}
\newcommand{\qicsinjref}{Q_{\gamma}^{\rm ics,R}}
\newcommand{\licslos}{L_{\rm e}^{\rm ics}}
\newcommand{\licslosref}{L_{\rm e}^{\rm ics,R}}
\newcommand{\qgginj}{Q_{e}^{\gamma\gamma}}
\newcommand{\qgginjref}{Q_{e}^{\gamma\gamma,R}}
\newcommand{\lgglos}{L_{\gamma}^{\gamma\gamma}}
\newcommand{\qannih}{Q_{\gamma}^{\rm ann}}
\newcommand{\lannih}{L_{e}^{\rm ann}}
\newcommand{\lssa}{L_{\gamma}^{\rm ssa}} 
\newcommand{\tthom}{\tau_{\rm T}}

\newcommand{\taumirror}{\tau_{\rm mir}}

\newcommand{\tcross}{t_{\rm cr}}
\newcommand{\zcrit}{z_{\rm crit}}
\newcommand{\tcrit}{t_{\rm crit}}

\def\gsim{\mathrel{\raise.5ex\hbox{$>$}\mkern-14mu
             \lower0.6ex\hbox{$\sim$}}}

\def\lsim{\mathrel{\raise.3ex\hbox{$<$}\mkern-14mu
             \lower0.6ex\hbox{$\sim$}}}

\shorttitle{GRB}
\shortauthors{Mastichiadis et al.}

\begin{document}


\title{The Supercritical Pile Model for GRBs}


\author{A. Mastichiadis}
\affil{Department of Physics, University of Athens, Panepistimiopolis,
  GR 15783, Zografos, Greece}
\author{D. Kazanas}
\affil{Laboratory for High Energy Astrophysics, NASA/GSFC, Code 661,
Greenbelt, MD 20771}



\begin{abstract}
We present the spectral and temporal radiative signatures expected 
within the ``Supercritical Pile" model of Gamma Ray Bursts (GRB).
This model is motivated by the need for a process that provides the 
dissipation necessary in GRB and presents a well defined scheme 
for converting the energy stored in the relativistic protons of the 
Relativistic Blast Waves (RBW) associated with GRB into radiation; 
at the same time it leads to spectra which exhibit a peak in the 
burst $\nu F_{\nu}$ distribution at an energy $E_p \simeq 1$ MeV in 
the observer's frame, in agreement with observation and largely 
{\sl independent} of the 
Lorentz factor $\Gamma$ of the associated relativistic outflow. Futhermore,
this scheme does not require (but does not preclude) acceleration of 
particles at the shock other than that provided by the isotropization 
of the flow bulk kinetic energy on the RBW frame. 
In  the present paper we model in detail the evolution of protons, 
electrons and photons from a RBW 
to produce detailed spectra of the prompt GRB phase  as a function of time from 
across a very broad range spanning roughly $4 \, {\rm log}_{10} \Gamma$ decades  
in frequency. The model spectra are in general agreement with  
observations and provide a means for the delineating of the model 
parameters through direct comparison with trends observed in 
GRB properties. 

\end{abstract}




\pagebreak
\section{Introduction}

The discovery of GRB afterglows by BeppoSAX (Costa et al. 1997) and the ensuing 
determination of their redshifts (van Paradijs et al. 1997) by and large settled 
the issue of their distance and luminosity. This discovery, then, settled 
also the issue of their energetics in favor of emission by Relativistic 
Blast Waves (RBW) moving toward the observer with Lorentz factors
$\Gamma \sim 100 - 1000$, as it had been discussed earlier by Rees \& 
M\'esz\'aros (1992) and M\'esz\'aros \& Rees (1992). The relativistic 
boosting of the radiation emitted at the rest frame of the RBW (by $\simeq 
\Gamma^4$ in luminosity) then
resolved also the issue of their huge energy 
budget requirements
(if at cosmological distances) and brought them to agreement with the 
energy release in stellar gravitational collapse. At the same time it
provided consistency between the huge implied GRB luminosities and their
apparently thin spectra (Krolik \& Pier 1991; Fenimore, Epstein \& Ho 1993; 
Barring \& Harding 1995). 
The same considerations provided also (Rees \& M\'esz\'aros 1992) 
an order of magnitude relation between the GRB duration $\Delta t$ and 
the size of the radiating region, namely $R \simeq
10^{16} (\Delta t/30 \; {\rm sec})\, (\Gamma / 100)^2$ cm. 

These estimates of the kinematic state of the GRB emitting plasma 
have been supplemented by certain dynamical
considerations. For example, following the work of Shemi \& Piran (1990) 
it has been generally accepted that a certain amount of baryons must 
be carried off with the blast waves responsible for the GRBs. This 
baryon contamination has even been deemed necessary for the efficient
transport of the GRB energy away from the environs of its ``inner engine", 
else the entire blast wave's internal energy would be converted into radiation 
on very short time scales, leading to events of very different temporal and
spectral appearance (e.g. Paczy\'nski 1986) than the observed GRB. 
In fact, the original models of Rees \& M\'esz\'aros (1992) and M\'esz\'aros 
\& Rees (1992) relied on the presence of a rather precise amount of 
baryons within the GRB fireball: enough to keep the fireball optically thick and
thus allow the conversion of its internal energy to directed motion
upon its expansion, but not too many as to render it only mildly (or 
even non-) relativistic. Even in the more recent (and perhaphs more 
plausible) variants of the same model that use Poynting  flux (rather 
than photon energy density) as the agent responsible for the RBW 
acceleration (see Vlahakis \& K\"onigl 2002, 2004), the circumstellar 
matter swept up by the RBW to the radius of $R \sim 10^{16}$ cm, contains
roughly as much energy stored in protons as in magnetic field. In either 
case, the models are called to shed light to these two generic, still open, 
issues of the GRB dynamics: (a.) The acceleration of the associated 
Relativistic Blast Waves to Lorentz factors $\Gamma \sim 10^2-10^3$ (b.) 
The dissipation of the energy stored in protons and/or magnetic fields at 
the rates necessary to produce the prompt GRB emission with the proper 
attributes. Both these issues are still open in GRB physics.

An altogether different issue associated with the prompt GRB 
emission is that of their spectra. The differential photon 
GRB spectra can be fit very well by the so-called Band-spectrum (Band 
et al. 1993) that consists (asymptotically) of
two power laws of indices $\alpha$ and $\beta$ joining smoothly
at a break energy $E_b$, i.e. 
\begin{equation}
N(E) \propto \left\{ \begin{array}{cc}
			E^{\alpha} \; e^{-(\alpha - \beta)E/E_b} & 
\mbox {for $E <  \, E_b$}\\
			E^{\beta} \; E_b^{\alpha - \beta} \; e^{-(\alpha - \beta)}  
& \mbox {for $E >  E_b$}
					\end{array}
  			\right.
\end{equation}
It was pointed out by Malozzi et al. (1995), and confirmed by a larger  
sample of {\sl BATSE} data (Preece et al. 2000), that the values of $E_b$ 
are narrowly distributed around $E_b = 200$ keV with a similarly narrow 
distribution for $\alpha$ around the value $\alpha =-1$, while the distribution 
of $\beta$  has a maximum near $\beta \simeq -2.3$ and extends to values 
$\beta \lsim -4$, with only few bursts having $\beta > -2.3$. These values imply 
that the GRB peak energy of their $\nu F_{\nu}$ spectra, (i.e. the peak energy 
of their emitted luminosity $E_{\rm p} = (\alpha + 2) \, E_b /(\alpha - \beta)$) 
is equally narrowly distributed about the same energy 
which, when corrected for the GRB redshift 
($z_{_{\rm GRB}} \sim 1-2$), shifts close to $E_{\rm p}\simeq 0.5$ MeV. 

Both the presence of the narrowly distributed energy $E_{\rm p}$ and
the value of the low energy index $\alpha \simeq -1$ are hard to 
understand within the conventional wisdom models which suggest that
the observed prompt GRB $\gamma-$ray emission is due to synchrotron
radiation by relativistic electrons. Under these assumptions, 
$E_p$ should be proportional to 
$\Gamma^4$ (two powers of $\Gamma$ come from the synchrotron emission, one
from the magnetic field - assuming equipartition with the postshock matter,
and one more from the transformation of this energy to the lab frame). 
This strong dependence on the value of $\Gamma$ would imply a rather unique 
value for this latter parameter, else the $E_{\rm p}$ range would be much broader
than observed, in disagreement with observation. While there may indeed
be a reason for a very narrow range in the values of $\Gamma$ consistent with
the observed range of $E_{\rm p}$, as of today such a reason is unknown (at 
least to the authors). If the same emission is due to electrons accelerated 
to energies beyond those implied by the shock Rankine - Hugoniot conditions
(i.e. $\gamma_e \simeq \Gamma$), or to a low energy cut-off $E_c$ in the 
electron injection spectrum (with $E_c \gg \Gamma \, m_ec^2$), 
the presence of a rather well defined peak in the GRB $\nu F_{\nu}$
spectra is even harder to understand, as in either case there is no
apparent reason for such a well defined energy scale.

Considerations along similar lines do argue for very specific values
for the low energy power law index $\alpha$: If the observed peak in
the $\nu F_{\nu}$ spectra is due to synchrotron emission by a 
$\delta-$function--like electron distribution, $\alpha$ should 
be that corresponding to the single electron synchrotron emissivity, 
$\alpha = - 2/3$  (Katz 1994a), while if the 
observed peak is due to a low energy cuf-off in the electron injection 
spectrum, the associated extremely fast cooling should lead to $\alpha =
-3/2$ (Ghisellini 2002). The observed values extend outside the above 
range, while the most likely values are inconsistent with either of these 
assumptions in the context of synchrotron radiation for the prompt GRB emission.

The discovery of GRB afterglows,  following the prediction by
Katz (1994b) shifted attention from the 
prompt GRB phase to that of the afterglow. Novel issues related to the 
physics of GRB ouflows have since emerged, e.g. the narrow range 
of the total GRB energy budget (Frail et al. 2001; Bloom et al. 2003) 
and the correlation of GRB luminosities with their spectral lags 
(Salmonson \& Galama 2003). In general, the afterglows extended the 
frequency range of GRB study to the X-rays, optical, IR and radio thus
greately expanding our means of probing of the RBW of GRB (see Zhang \& 
M\'esz\'aros 2004; Piran 2004 for reviews). In the midst of this new
flurry of activity, the issues of dissipation of the GRB proton 
kinetic energy and the narrow $E_{\rm p}$ distribution, while quietly ignored, 
have remained largely unanswered. One of the few attempts to address these 
issues was that of Kazanas, Georganopoulos \& Mastichiadis (2002; hereafter 
KGM): In search of a well defined and tractable dissipation mechansim, 
they proposed a process that utilises proton - photon collisions to convert
the energy stored in protons behind the forward shock of the RBW to electrons 
(and then into radiation). KGM provided the kinematic and dynamic 
thresholds for this process to take place and showed that, in addition, it 
produced spectra that peaked at several well defined energies, 
namely at $ \sim \;2 m_ec^2/\Gamma^2, \; 2 m_ec^2, \; m_ec^2 \Gamma^2$, 
 {\sl in the observer's frame}. 
The interesting feature of this model is  that the combination of the 
threshold for pair production energy and the final Lorentz boost to 
the observer's frame leads to a spectral peak at $\sim m_ec^2$  
{\sl largely independent} of the Lorentz factor $\Gamma$ of the RBW, 
provided that the latter is larger than a threshold value which depends 
on the value of the magentic field of the RBW. 
Therefore, the process described by KGM provides a framework for resolving both the 
proton-to-electron-energy-transfer and the narrow $E_p$--range problems 
in a single fell swoop. 
Besides the above properties, this model differs from most in the 
present literature in several respects: (a) It does not require (but does
not preclude either) the presence of accelerated particle populations, other
than those produced by the isotropization of the RWB kinetic energy behind the
shock. (b) It does not require equipartition between the proton and 
electron enery densities. (c) It posits that the observed radiation in the 
X - $\gamma$ ray band is upscattered (and then blue-shifted) synchrotron 
radiation rather than simply blue-shifted synchrotron radiation.

The purpose of the present paper is to present detailed calculations based
on the model outlined in KGM, which for reasons explained below 
(and in the original reference), is referred to as the ``Supercritical 
Pile" model for GRB. In \S 2 of 
this note we provide a qualitative description of the basic notions that 
underlie the ``Supercritical Pile" model and how they apply to GRB. 
In \S 3 we describe the details of the model and of the numerical method 
employed for the solution. In \S 4 we describe the numerical tests used 
in making certain that our results make sense and the code works as planned
in the case that contains no electrons but produces all necessary electrons
from the proton radiative instability. In \S 5 we repeat the calculations
of \S 4 in the more realistic case that includes also the effects of 
the presence of electrons. 
We also present light curves and spectra produced within the model for 
different values of the relevant parameters. Finally in \S 6 the resuts are
summarised and conclusions are drawn.

\section{Dissipation: The Proton-Pair-Synchrotron (PPS) Network}

The name of our model derives from the (at first sight incredulous) similarity
between the RBW of a GRB and a nuclear pile. However, a closer inspection
- albeit with our model in mind - reveals the following similarities: 
(a) They both contain a large amount of free energy, stored in relativistic 
protons in the GRB case and in nuclear binding energy 
in the case of the pile. 
(b) This energy can be released explosively - i.e. on the time scale of 
crossing the relevant size by photons or neutrons respectively, once identical 
criticality conditions are met. Furthermore, besides the possibility of the 
explosive release of the energy contained in 
the relativistic protons of a RBW, our model also provides an account
for the observed peak in the $\nu F_{\nu}$ GRB spectra; interestingly 
it achieves this in one of its variants conceived to provide agreement 
between the numerical values of the parameters that determine the 
criticality condition with those associated with  the GRB settings.
It is interesting to note that processes involoving the exponential increase
of the number of photons or their luminosity akin to the criticality condition
of a nuclear pile discussed above and in the rest of this paper are not 
new in astrophysics; we are aware of two such instances, one involving the energy
released through the comptonization of soft photons by hot electrons (Katz 1976) 
while the other the equilibration of plasmas through the Double Compton process
(Lightman 1981).

The process we model in detail in the following has two distinct aspects,
along the lines of the two GRB issues discussed above. The first one 
concerns the transfer of the energy stored in a population of relativistic 
hadrons (protons) into leptons. 
This has been discussed originally by Kirk \& Mastichiadis (1992; KM92)
and modeled in more detail in Mastichiadis \& Kirk (1995; MK95) 
and Mastichiadis, Protheroe \& Kirk (2005; MPK05). This same process 
was then modified to include the effects of relativistic motion of
the plasma in the possibility that some of its photons could be scattered 
by matter (a ``mirror") located upstream along the direction of motion. 
This leads to a relaxation of the threshold conditions obtained in KM92 and was
discussed by Kazanas \& Mastichiadis (1999; KM99). The second aspect concerns
the spectra that result from the combination of these processes as was 
originally discussed in KGM and their relation to the energetics of GRB.

To formulate the conversion of proton energy to leptons through the $p \, 
\gamma \rightarrow p\, e^+ e^-$ process one assumes the presence of a population 
of relativistic protons of the
form  $N(\gamma_p) = n_0 \, \gamma_p^{-\beta}$; 
the requisite photons are provided by the synchrotron radiation  of the 
pairs produced in the proton - photon intereactions, 
with these two processes combined into a reaction network. 
This network involves both kinematic and dynamic thresholds.
The physical arguments that provide these thresholds are rather simple: 
There is a {\sl kinematic}
threshold for the reaction $p \, \gamma \rightarrow p \,e^+ e^-$; for 
a photon of energy $E_{\gamma}$, there is a critical value 
$\gamma_c = 2 m_ec^2/E_{\gamma}$ of the proton Lorentz factor 
for the reaction to take place (assuming a head-on photon-proton 
collision, 
else this condition should involve the cosine of the angle between these
particles), with each lepton  produced by this reaction 
with energy $\gamma_c \; m_ec^2$ (on the average). The reaction network 
will be {\sl self-contained} if these pairs can produce, through 
the synchrotron process, the photons needed to effect the pair 
production, i.e. if $E_{\gamma} = E_s = b \, \gamma_c^2\, 
m_ec^2$ ($b$ is the magnetic field $B$ of the plasma in units of the 
critical magnetic field $B_c = m_e^2 \, c^3/c \hbar$ = $4.4 \; 10^{13}$ 
G). This leads to the following kinematic threshold for the reaction 
network $\gamma_c^3 \, b \simeq 2$ or $\gamma \simeq (2/b)^{1/3}$, as 
discussed in KM92.

The combined network of the photon and pair producing reactions will 
also be {\sl self -sustained} if {\sl at least one} of the synchrotron 
photons produced by the $e^+ \, e^-$ pairs produces another pair before 
escaping the volume of the plasma in a reaction with a sufficiently 
energetic proton (i.e. one that fulfills the kinematic threshold). This 
results in a condition for the column density of the plasma which is 
identical to that of a critical nuclear chain reaction (we re-iterate 
for the benefit of the reader that criticality is a condition
on the column density rather than the - erroneously 
referred to in the popular literature - mass of the pile). Therefore, 
the plasma column density (at the proton critical energy), must be 
greater than the inverse
of the number of synchrotron photons emitted by the electrons of 
Lorentz factor $\gamma_c$. The number of synchrotron photons  
just above the 
thershold energy is ${\cal N} = \gamma_c/b\, \gamma_c^2 = 1/b \, 
\gamma_c$ leading to the dynamic condition $\tau_{p \gamma}
\simeq \sigma_{p \gamma} \, R \, N_p(\gamma_c) \, \gamma_c \gsim
b \, \gamma_c$. Taking into account the kinematic threshold condition 
this reads $\sigma_{p \gamma} \, R \, n_0 \gsim b^{(1 - \beta/3)}$,
in agreement with the result of KM92.

Mastichiadis \& Kirk (1995) and Mastichiadis, Protheroe \& Kirk (2005)
have explored the above process numerically in 
detail and found the semianalytic estimates above to be in excellent 
agreement with the more detailed numerical studies. The instability 
depends on both thresholds as described and converts the energy density
stored in relativistic protons (those with energy above that of the 
threshold $\gamma_p \ge \gamma_c$) into 
pairs on the plasma crossing time scale $R/c$. During the initial (linear) 
regime of the instability the pair and photon numbers increase exponentially
($\propto e^{st}$) with the exponent $s$ being greater the larger the value
of the proton density $n_0$ (i.e. the farther the network is above 
threshold), while $s=0$ at precisely the threshold value
of this parameter.

The RBW associated with the GRB provide a natural setting for the production 
of relativistic proton populations, thus lending them as natural sites
where the above considerations should apply. However, relativistic motion
alone does not in general change the kinematic and dynamic threshold
conditions obtained above. It is conceivable that the presence of a 
relativistic proton population that extends to sufficiently high energies 
could satisfy the kinematic threshold of the instability; however, the 
``standard" GRB model parameters yield  column densities for the 
relativistic protons (these are just those in the matter ``swept-up" 
by the RBW) far smaller than necessary to satisfy the dynamical 
threshold condition. 

As pointed out in Kazanas \& Mastichiadis (1999; KM99) the situation
can change drastically in the presence of matter along the direction of 
relativistic motion which could isotropize through scattering the radiation 
emitted by the relativistically moving plasma. In this case, the RBW can 
``catch-up" with the scattered photons which, when re-intercepted by the RBW, 
 appear on its rest frame blue-shifted by a factor 
$4\Gamma^2$. If those were the synchrotron photons of energy $\epsilon_s
\simeq b \, \gamma^2$ they will now have an energy $\epsilon'_s \simeq
4b \, \gamma^2  \Gamma^2$ (in units of $m_ec^2$). Therefore, the 
kinematic condition for the 
{\sl reflected} photons to produce $e^+ e^-$-pairs gets modified to
\eqb 
\gamma \, \epsilon'_s \gsim 2 ~~{\rm or}~~
2b \, \gamma^3 \, \Gamma^2 \gsim 1. 
\eqe
Making the further restrictive 
assumption that the only relativistic protons are those  
which are produced
by the isotropization of the swept-up matter by the RBW, we can set
$\gamma = \Gamma$ to obtain the kinematic threshold condition
\eqb
\Gamma^5  b \gsim 2 ~~{\rm or} ~~ \Gamma \gsim 
\left(\frac{1}{2b}\right)^{1/5}.
\label{kthresh} 
\eqe

One can employ the same considerations used above to 
obtain the dynamical threshold of the instability in this case too.
The number of photons produced by an electron of Lorentz factor
$\Gamma$ on the RBW will now be ${\cal N} \simeq 
\Gamma/b \, \Gamma^2 = 
1/b \, \Gamma$;  assuming that a fraction $\tau_{\rm mir}$  of these photons are 
scattered by the mirror (and are therefore re-intercepted by the RBW) leads to
\eqb
\sigma_{p \gamma} \, R \, n \gsim \frac{b\Gamma}{\tau_{\rm mir}} ~~{\rm or}~~
2\tau_{\rm mir} \, \sigma_{p \gamma} \, R \, n \, \Gamma^4 \gsim 1
\label{dthresh}
\eqe
with the latter expression obtained from the former with the 
use of the kinematic threshold condition (we would like to thank P.
M\'esz\'aros for pointing our an error in an earlier version of 
the above expression). 
Here $\sigma_{p \gamma}$ is the photopair (Bethe-Heitler)
cross section as given by Motz et al. (1969).
The quantity $n$ above 
now denotes the value of the ambient density (one should 
note that the column density is a Lorentz invariant so that its value
as given above is identical to its value on the RBW rest frame).
This condition can be satisfied for  values of the GRB
parameters $n \simeq 10^3 \, n_3$ cm$^{-3}$ and $R = 10^{16} \, R_{16}$ cm
for values of $\Gamma \gsim 100 \, (n_3 R_{16})^{-1/4}$ (assuming 
$\tau_{\rm mir}
\simeq 1$; see also discussion in \S 6).

\section{Particle evolution and radiation: The kinetic equation approach}

Having outlined the qualitative aspects of the `Supercritial Pile' model 
we proceed to describe in detail the procedure used in simulating this model.
We consider a Relativistic Blast Wave (RBW) 
moving with speed $\upsilon_{0}=\beta_\Gamma c$,
where $\beta_{\Gamma} =(1-\Gamma^{-2})^{-1/2}$ and $\Gamma$ the
bulk Lorentz factor of the flow which we assume to
be constant. It has a radius $\Rbrw(t)$
as measured from the origin of our coordinate system (assumed to 
be the center of the original explosion) and
it contains  protons and electrons.
If there is no particle acceleration taking place
we can, conservatively, assume that both species have been
isotropized and that in the flow rest frame they have  distributions
with average Lorentz factors $\langle \gamma_p \rangle \simeq \langle 
\gamma_e \rangle \simeq \Gamma$. 
Under these conditions, protons carry a significant fraction of the
blast wave energy. 
The electron and proton distributions will evolve in time on the 
RBW frame since both species will suffer from various types
of energy losses, physical escape, etc. At the same time they
will radiate by synchrotron, inverse Compton and possibly other
processes. 

We assume also the presence of scattering material 
(which we will call the ``mirror")
ahead of the advancing shock. This is located between
radii $R_{s,1}$ and $R_{s,2}$ and it is assumed to have
a uniform electron density $\nelsc$ and a corresponding
Thomson optical depth 
$\taumirror = \nelsc\sigmaT (R_{s,2}-R_{s,1})$ (at present, 
we consider only Thomson scattering as the main process that
deflects the RBW synchrotron photons; other possibly important 
processes are at this stage ignored). Therefore part of the 
radiation emitted from the particles in the RBW 
will be intercepted by this mirror and scattered back,
reentering the shock at a later time when this will 
have moved to a different location. The reflected (and amplified in 
energy) photons will contribute to the particles' (protons and 
electrons) energy losses, an additional feature of the 
present problem.

To treat the radiative transfer we assume that the relativistic
blast wave is expanding in a spherical fashion, however, due 
to relativistic beaming an observer receives the radiation coming mainly
from a small section of it of lateral width $\Rbrw/\Gamma$ and 
longitudinal width $\Rbrw/\Gamma^2$ in the lab frame but $\Rbrw/\Gamma$ 
on the flow frame. Therefore, at the RBW frame, one can approximate
the emitting region by a spherical `blob' of radius $R_b=\Rbrw/\Gamma$ 
and solve the radiative transfer problem for that region. The particles 
in the other segments of the blast wave may follow a similar evolution, 
however the radiation coming from them will not be observed at the Earth
as it is beamed away from us at angles $>\Gamma^{-1}$, neither will be 
intercepted, for the same reason, by the segment in consideration and 
it will not affect the evolution of the particle distribution in it. 

Alternatively we can assume, instead of  
a sphere, a segment of opening angle $\Theta$.
Our analysis does not change apart from the fact
that the Doppler factor 
$\delta=[\Gamma(1-\beta cos\theta)]^{-1}$
should be introduced when making the transformation
of the radiation patterns betwwen the RBW
(comoving) and Earth (stationary) frames. 

The problem of particle evolution in a homogeneous source
region containing protons, electrons and photons was
formulated and solved numerically by Mastichiadis \& Kirk 
(1995) and more recently by Mastichiadis, Protheroe \& Kirk
(2005)
adopting the kinetic equation approach.
Herein we follow the same method, making the changes
which are appropriate for the present case. 
The equations to be solved can be written in generic form
\eqb
{{\partial n_i}\over{\partial t}} + L_i +Q_i=0 
\eqe
where the index $i$ can be any one of the subscripts `p', `e' 
or `$\gamma$' referring to protons, electrons or photons respectively.
The operators $L_i$  denote losses and escape from the system  
while $Q_i$ denote injection and source terms. These will be
defined further below.

The unknown functions $n_i$ are the differential
number densities of the three species and
the physical processes to be included in the kinetic equations are:
\begin{enumerate}  
\item
Proton-photon (Bethe-Heitler) pair production which acts as a loss term
for the protons ($\protloss$) and an injection term for the electrons
($\qbhinj$)
\item
Photopion production which also acts as a loss term for the protons
($\protlopi$) and and an injection term for both electrons
($\elinpi$) and photons ($\phinpi$).
\item
Synchrotron radiation which acts as an energy loss term for electrons 
($\lsynlos$)
and  as a source term for photons ($\qsyninj$)
\item
Synchrotron self absorption which acts as an absorption term 
for photons ($\lssa$)
\item
Inverse Compton scattering (in both the Thomson and Klein-Nishina regimes)
which acts as an energy loss term for electrons ($\licslos$)
and as a source term for high energy photons
and a loss term for low energy photons, both effects
included in  $\qicsinj$
\item
Photon-photon pair production
which acts as an injection term for electrons ($\qgginj$)
and as an absorption term for photons ($\lgglos$)
\item
Electron-positron annihilation
which acts as a sink term for electrons ($\lannih$)
and as a source term for photons ($\qannih$)
\item
Compton scattering of radiation on the cool pairs, which impede
the free escape of photons
from the system. 
\end{enumerate}

A feature that differentiates our present study from MK95
and MPK05
is that processes involving photons in the left hand side
(i.e. processes 1, 2, 5 and 6) can take place either with
photons produced directly or with the photons
which have been reflected by the mirror assumed to exist
in front of the advancing shock.

In order to calculate the reflected photon number density
as seen in the rest frame of the blob we use Eqn. (2)
of B\"ottcher and Dermer (1998)
corrected for Klein-Nishina effects.
The reflected photon number density 
when the blob is at some distance $z$
from the center is given by 
\eqb
\nphotre(s_2,z_2)={{3\nelsc\sigmaT}\over{8c}}
\int_{\mu_2^{min}}^{\mu_2^{max}}d\mu_2
\int_{x_2^{min}}^{x_2^{max}} dx_2 D_1^2
{{\dot N(s_1,\Omega_1;z_1)}\over{x_1^2}}
(1+cos^2\chi)
\eqe
where  
$s_1$ and $s_2$ are the photon energies
before and after reflection,
$z_1$ and $z_2$ are the RBW spatial 
coordinate along the direction of 
motion at the instance of photon emission
and reception,
$x_1$ and $x_2$ are the distances of the reflection
point from the RBW while this is at distances $z_1$ and
$z_2$ respectively,
$\mu_1$ and $\mu_2$ are the cosines of the angles that
$x_1$ and $x_2$ make with the axis $z$, while
$\chi$ is the scattering angle in the RBW. The energies
of the photons before and after reflection are related by  
$s_1=s_2/(D_1D_2)$,  where
$D_1=\left[\Gamma(1-\beta_\Gamma\mu_1)\right]^{-1}$,
$D_2=\Gamma(1-\beta_\Gamma\mu_2)$.
The light travel-time effects are of 
great importance in our calculations. The causality condition 
\eqb
z_1 = z_2 - \beta_\Gamma(x_1+x_2)
\eqe
along with the condition $z_1>0$ 
implies that the blob does not receive any reflected photons
for as long as it is at
\eqb
z < \zcrit = R_{s,1}{{2\beta_\Gamma}\over{1+\beta_\Gamma}}.
\eqe

Therefore, if we assume for simplicity, that
the RBW has reached a constant Lorentz factor
$\Gamma$ when it is some distance $z_0$ from the origin,
then we can relate the current position of
the RBW, $z$, with the time elapsed since then  
 by the relation
\eqb
z = z_0 + \beta_\Gamma c t.
\eqe
With this hypothesis 
we can define a time 
\eqb
\tcrit={\zcrit\over{\beta_\Gamma c}}
\eqe
with the property that  only when $t>\tcrit$, 
photons produced at earlier times will be
reentering, due to reflection, the RBW.

A second point is 
that from the processes listed above photopion production,
electron-positron annihilation,
and Compton downscattering
turn out to be totally unimportant for the  
parameters of the cases we will examine here. 
Photon-photon pair
production also turns out to be of marginal importance.
So, despite the fact that we keep all the aforementioned
processes in the code, our focus is on the processes that 
provide the driving terms in the proton loop, i.e. the Bethe 
Heitler pair production (the source of pairs) [1], the electron synchrotron 
(the source of photons)[4] and inverse Compton scattering (the 
photon energization process)[5].
Further discussion on these points will be given in the
last section. 

With the inclusion of the leading terms only, the kinetic 
equations for each species read

\begin{itemize}
\item
Protons
\eqb
{{\partial\nprot}\over{\partial t}} + \protloss 
+ H(t-\tcrit)\protlossref = 0
\label{protkinet}
\eqe
\item
Electrons
\eqb
{{\partial\nelec}\over{\partial t}} + \lsynlos + \licslos
+ H(t-\tcrit)\licslosref 
=  \qbhinj + H(t-\tcrit)\qbhinjref +  \qgginj + H(t-\tcrit)\qgginjref
\label{eleckinet}
\eqe
\item
Photons
\eqb
{{\partial\nphot}\over{\partial t}} + {\nphot\over\tcross} +
\lgglos + \lssa =\qsyninj + \qicsinj + H(t-\tcrit)\qicsinjref
\label{photkinet}
\eqe
\end{itemize}

We repeat here some of the comments made in MPK05 
regarding these equations:

1. When the various terms above are written explicitly,
equations (\ref{protkinet}),~(\ref{eleckinet}) and (\ref{photkinet})
form a non-linear system of coupled partial  integrodifferential
equations which are solved numerically.

2. The various rates are written in a  manner that conserves power
in the exchanges between the species.

3. Without the proton and reflection terms 
the system becomes identical to those
considered in the `one-zone' time-dependent
leptonic models -- see for example Mastichiadis \& Kirk (1997).

We note however that the system of the equations is
not identical to the one solved in MPK05:

4. The reflected photon number density which is computed by means of 
Equation (2) of B\"ottcher \& Dermer (1998) depends on $\nphot(x,t')$ 
at all previous times (i.e. $t'<t$).
Thus no extra equation is required for this component.
The function $H(t-\tcrit)$
is the Heaviside function denoting that the reflected
photons start playing a role for $t>\tcrit$, i.e. when
the position of the RBW is inside the reflected photon zone
--see Eqn. (6).
Therefore at this point retardation effects become important.

\subsection {Computational Details}

The unknown functions $n_i$ in Eqns (5)-(7)
are the differential
number densities of the three species, 
which are normalised as follows:
\eqb
\textrm{Protons:}&\tilde\nprot (\gammap,t)d\gammap\,=\,\sigmaT
R\nprot(E_p,t) dE_p
&\textrm{with }
\gammap\,=\,{{E_p}\over{\mprot c^2}}
\\  
\textrm{Electrons:}
&\tilde \nelec(\gammae,t)d\gammae=\sigmaT R\nelec(E_e,t) dE_e&
\textrm{with } \gammae\,=\,{{E_e}\over{\melec c^2}}
\\
\textrm{Photons:}
&\tilde \nphot(x,t)dx\,=\,\sigmaT R\nphot(\epsilon_\gamma,t)
d\epsilon_\gamma&
\textrm{with } x\,=\,{{\epsilon_\gamma}\over{\melec c^2}}
\eqe
and the time $t$ has been normalised in all equations to the light-crossing
time of the source $\tcross=R_b/c$.  

With this definition
the photon compactness becomes (we drop now the tilde for
convenience)
\eqb
\lphot&=&{{1}\over{3}}\int dx x {\nphot(x,t)\over\tcross}
\label{photcompact}
\eqe
while the compactness of the reflected photons can be defined
in an analogous fashion
\eqb
\lphotref&=&{{1}\over{3}}\int dx x {\nphotref(x,t)\over\tcross}.
\label{photcompact}
\eqe
Also the Thomson optical depth of the cooled electrons
inside the blob is given by
\eqb
\tthom(t)&=&\int d\gamma~\nelec(\gamma=1,t)
\eqe
Finally, we can define the magnetic 
compactness
\eqb
\bcomp&=&\sigmaT R {{B^2}\over {8\pi\melec c^2}}.
\eqe
For $\bcomp > max(\lphot,\lphotref)$ the electron cooling
is controlled by synchrotron while in the opposite case
by inverse Compton scattering.

We note that the treatment of synchrotron and inverse Compton scattering 
has been improved over that described by MK95 in that the full
emissivities are incorporated, rather than delta-function approximations.
On the other hand, the Bethe-Heitler pair production still uses 
the delta-function approximation for the resulting electron-positron
spectra. However, as shown in Mastichiadis, Protheroe, Kirk (2005; MKP05
), this approximation does not alter the main results of KM92, MK95 or 
those of the present study.

\section{Modeling the PPS Network - No Electrons Present}

Numerical studies of the proton loops in the absence of reflection have 
been presented already in MK95 and MPK05. For the clearer understanding of
the effects and modifications caused by the presence of reflection
we present in this section a similar numerical
study taking reflection into account.
As discussed in KM92, MPK05 and in 
Section 2 of the present paper, a relativistic proton distribution 
becomes unstable to the Proton-Pair-Synchrotron (henceforth PPS) 
loop if the two threshold conditions outlined there are 
satisfied (see relations (\ref{kthresh}, \ref{dthresh}) of Section 2).
This instability manifests itself with a spontaneous photon-pair growth 
characterised by an exponential rise in their respective densities of the 
form $\nphot\sim\nelec\sim e^{st}$, with $s$ being a function of both 
the number density and maximum energy of the protons. A second 
characteristic is that the maximum spectral luminosity of the produced 
photons occurs, at least during the initial stages of the instability 
growth, at the critical synchrotron frequency of the produced electrons. 
In this section we perform a numerical analysis using the code described 
in section 3 to explore the various characteristics of the PPS+reflection 
(henceforth PPSR) loop. In this approach we shall ignore, for the moment, 
the presence of any electrons in our initial conditions and focus only 
on the protons. 

We consider, thus, that the RBW has a radius $\Rbrw=3.10^{16}$~cm $= R_{s,1}$
while it is cruising with a constant value of $\Gamma=10^{2.6}\simeq 400$. 
According to the arguments presented in Section 3, with a good approximation
one can consider the radiative transfer problem for only a segment of the RBW
which we take it to be, for convenience, spherical 
with a radius $R_b=\Rbrw/\Gamma=7.5~10^{13}$~cm. We assume that
in this spherical blob there is a monoenergetic proton
distribution with Lorentz factor $\gamma_p=\Gamma=10^{2.6}$ and
a number density $n_p=10^4~\rm{cm}^{-3}$. In the blob there is also a 
magnetic field of strength $B=44$ Gauss, which was chosen arbitrarily,
but in a way as to satisfy the thredhold condition (Eqn. 2). 
We also assume that ahead of the expanding RBW there is a ``mirror" of 
scattering material with total optical thickness $\taumirror=1$, 
distributed uniformely over its entire thickness. We apply the numerical 
code in a continuous run covering two distinct time intervals: (1) when 
the blob is still out of the reflection region ($t\le\tcrit$--Zone I)
and (2) after it has entered this region ($t>\tcrit$--Zone II). The 
beginning of the run ($t=0$) is set by the arbitrary requirement that 
10 blob crossing times (as measured in the RBW frame) will elapse before 
the RBW reaches $\zcrit$, the location where the reflected photons start 
entering the blob. Since we have restricted these set of  runs to initial 
conditions for which there are initially no photons or energetic electrons
within the plasma volume we consider, the time history of the blob in Zone I 
is irrelevant.

{\sl (i) The Growth Phase}

As long as the RBW is in Zone I the protons do not suffer
any substantial losses; therefore at the time the blob reaches $\zcrit$ 
it contains the original proton distribution while, according to our 
present assumptions, there are virtually no electrons or any other
sources of photons. However, when it enters Zone II, depending on the 
proton parameters, the PPSR loop will begin operating. The parameters 
given above were chosen in such a way as to fulfill the required 
conditions. Figure 1 shows the spectrum at five consecutive crossing 
times as seen by an observer at rest in the RBW frame. He/she will 
observe the directly produced photons to grow with a spectrum which is
peaked at an energy of about $\epsilon_{_S}\simeq b\Gamma^2=1.6~10^{-7}$ 
($m_ec^2$ units), the characteristic synchrotron energy of electrons of
Lorentz factor $\gamma_e=398$ in a magnetic field of (normalized) strength
$b=B/\bcrit=10^{-12}$ (for a discussion of these normalised units 
see Mastichiadis 2002). The photons in this case grow exponentially 
with index $s\simeq 5.1$. The reflected photons (represented by the dotted 
lines in this Figure) will be perceived in the RBW frame as having a 
much higher energy density than the directly produced internal photons, 
as they are amplified approximately by the factor $\alpha\Gamma^3$, 
where $\alpha$ is the albedo, while the maximum of their emission
will occur at $\epsilon_{_{\cal R}} \sim \Gamma^2\epsilon_{_S}\simeq 
b\Gamma^4\simeq 3~10^{-2}~m_ec^2$. The reflected photons will cause 
additional cooling of the electrons by inverse Compton scattering 
as the magnetic compactness $\bcomp$ remains constant while the reflected  
photon compactness ($\lphotref$) increases with time -- for all practical 
purposes cooling on the internally produced photons can be ignored
as $\lphot \ll \lphotref$). As a result of the inverse Compton cooling, 
the produced pairs will radiate an increasing fraction of their energy 
by this mechanism. 
Thus, the photon spectrum will consist of two components, the synchroton 
one at low energies and the ICS component at higher ones.
As the interactions between the hot electrons and reflected photons will 
occur mostly in the Klein-Nishina regime (since $\gamma_e
\epsilon_{_{\cal R}}\sim b\Gamma^5\simeq 10~m_ec^2$), the 
peak of this component 
will occur close to $\Gamma$. It is interesting to note that while 
the synchrotron component rises as $e^{st}$, the IC component rises like 
$e^{2st}$ (since it depends on electons and photons which both rise like 
$e^{st}$). In this respect this process resembles the SSC process, with 
the crucial differences that in our present case the target photons 
are not the internally produced synchrotron photons but the reflected 
ones and the prime movers are protons, not electrons. The behaviour 
described can be seen in Fig. 1. Note that the IC component is not 
reflected at the ``mirror" as the photons of this component appear 
in the mirror rest frame with energies that are well within the KN 
regime and therefore kinematic effects 
do not allow the reflection of this component.
\begin{figure*}[hbt]
\centerline{\epsfig{file=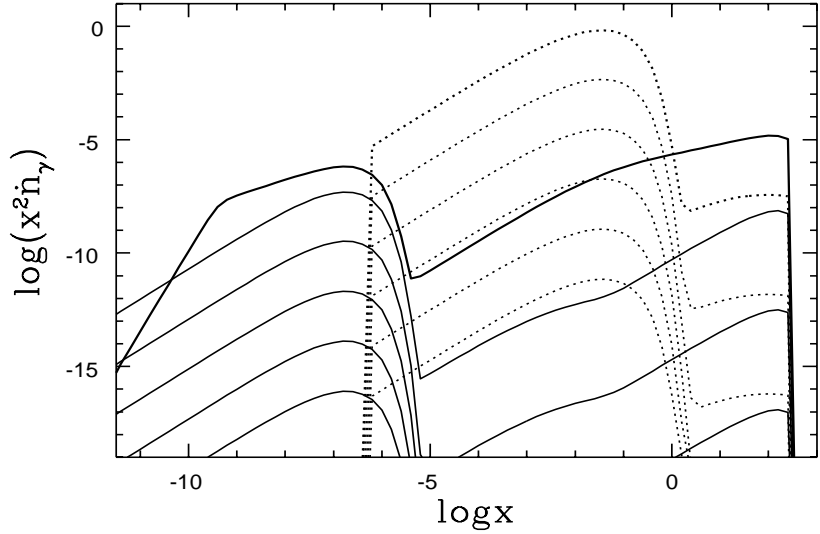,height=12.0cm,width=13.0cm}}
\caption{Snapshots, at consecutive crossing times,
 of the photon spectra (in code normalised units)  
during outgrowth as appear
in the RBW frame. Here $\Gamma=400$,
$n_p=10^4~\rm{cm}^{-3}$, B=44 G (comoving values)
and  $\tau_{\rm mir}=1$.
Photon energy is expressed in $m_ec^2$ units. 
Directly produced spectra are shown by full lines 
while the reflected spectra are shown by dotted lines.
The thick black line corresponds to the spectrum at the
peak of the directly
produced luminosity -- see Fig.2.}
\end{figure*}

{\sl (ii) The Saturation/Decay Phase}

The description given above concerns the early phases of
the pair/photon growth before their number has been
built substantially inside the system. When this eventually
happens there are two competing effects which become dominant.

a) The relativistic protons become increasingly shielded 
by the produced electron/positron pairs, with 
the reflected photons having an increasing
probability of undergoing  Compton scattering instead
of a pair-producing Bethe-Heitler collision. We note that the
rate of the latter reaction for a stationary proton distribution
is proportional to ${\cal{R}}_{BH}\propto \nprot\nphot
\propto e^{st}$, while the 
rate of the former is ${\cal{R}}_{ics}\propto \nelec\nphot
\propto e^{2st}$. Even if we were to assume that protons do not lose
energy, this effect alone would eventually  lead to saturation
of the instability.

b) The increasing number of photons above the critical
frequency of the PPSR loop and the ensuing production of pairs
at the expense of the proton energy leads to proton ``cooling".
Thus, while the total proton number does not change, their maximum
energy decreases below the threshold energy necessary for the 
loop to operate. As a result, no new pairs are injected into
the plasma which evolves by the cooling of the pairs already 
present and the decrease of its luminosity.

In reality both effects take place simultaneously.
However, if the loop operates close to threshold,
the proton losses are rather slow and the luminosity
decreases at a correspondingly slow  rate. On the other 
hand, operation of the loop well inside the unstable regime
will lead to much faster proton losses and consequently
faster photon cooling.

Figure 2  depicts the evolution of
the photon compactness (both directly produced and reflected)
as measured in the RBW frame once in Zone II.
We note that initially the photons grow with an index $s=5.1$
up to the point that $\lphotref>\bcomp$. At this point
the dominant electron cooling process shifts from synchrotron 
to inverse Compton and as a result the growth rate changes.
The photon density of the system saturates once it reaches a 
sufficiently  high level due to the combination of the
proton energy losses that
prevent further pair production and Compton saturation --
see above. 
In the particular example protons
lose about 1.5\% of their energy which is turned into
radiation. As we will see in the next section, depending 
on the initial parameters, protons 
can lose a substantial portion of their energy.

An important quantity in our analysis is the number density 
of cooled pairs. These are the pairs that are produced by the 
Bethe-Heitler pair production process, initially at Lorentz 
factors $\gamma=\Gamma$, and have subsequently cooled down to 
$\gamma \simeq 1$  by synchrotron and 
inverse Compton losses. The code described in Section 3
follows both the production and evolution of these 
particles and thus it can provide at each instant the 
quantity $n_e^{cool}(t)$ which, using the normalizations 
of Section 3.1, is equal to $\tau_T(t)$. As it can be seen 
from Fig. 2 their density builds quickly during the growth 
phase and after the onset of the saturation/loss phase it 
remains almost constant in the system. We also note that 
$\tau_{\rm T}(t)\ll 1$, i.e. in our model ( and for the
chosen values of the parameters) the RBW continues 
to be optically thin throughout the burst and hence its directly 
produced spectrum is not modified by repeated Compton 
scatterings. This is different from the results of MPK05 
who investigated the properties of the PPS loop in the 
absence of reflection. The reason for this 
difference, as we have emphasized in Section 2, is the 
presence of upstream reflection which greatly relaxes the
threshold criteria and as a result both the required initial proton 
number density and the number of the produced pairs decrease.  

\begin{figure*}[hbt]
\centerline{\epsfig{file=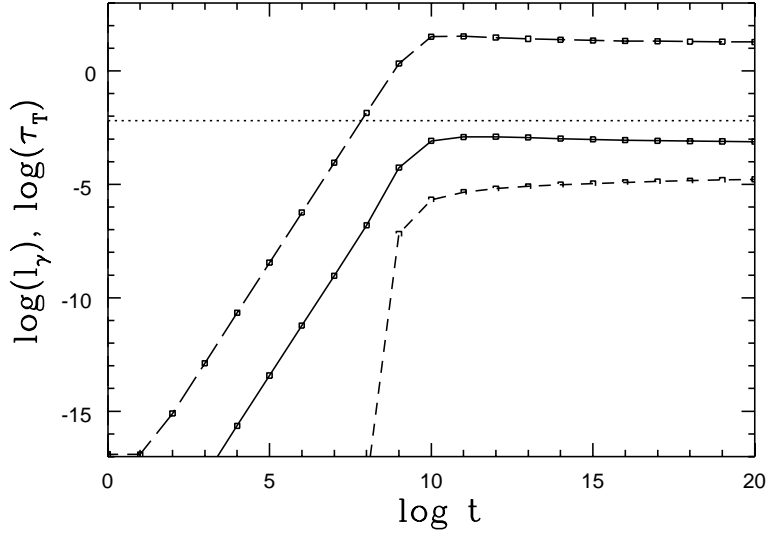,height=12.0cm,width=12cm}}
\caption{Lightcurves of the directly produced
photon compactness (full line),
reflected photon compactness 
(long dashed line) and number density of cool pairs (normalised
here to the Thomson optical depth -- short dashed line)
as measured from the time
the RBW has entered the reflection zone 
for the parameters of the previous figure.
Time is measured
in blob crossing times. The horizontal dotted line
denotes the magnetic compactness}  
\end{figure*}

{\sl (iii) The Observed Spectrum}

The processes described in the above subsections refer to quantities 
(electron and photon densities) as measured in the RBW frame. An 
observer at Earth will  observe the two spectral components 
(synchrotron and inverse Compton) shifted by $\delta$ in energy 
and by  $\delta^3$ in flux (in the case of a spherically symmetric 
expansion we can simply set $\delta=\Gamma$). This observer will not
see the reflected component as this is directed away from him/her. 
However, any reflected photon which is scattered on the cooled pairs
of the RBW will be isotropized in the frame of the flow and, upon 
transformation to the observer's frame, it will appear beamed within 
a cone of opening angle $\simeq 1/\Gamma$, just like the SSC components
with the same flux amplification ($\propto \delta^3$).
Thus the observed spectrum will have a third component, which we 
shall refer to as the ``doubly reflected" component and which is of special 
interest in our model as its peak emission lies in the 0.5 to 10 MeV 
regime for a wide set of parameters -- more details will be given in 
the next section. 

The most conservative case concerning the flux of this ``doubly 
reflected" component is that in which there are 
initially no ``cold" pairs in the RBW; however, as the PPSR 
loop begins to operate, an increasing number of them accummulates 
near $\gamma \simeq 1$ (see Fig. 2) by the cooling of those 
produced by the Bethe-Heitler process. Thus the relative 
importance of the doubly reflected component to the directly 
produced synchrotron one will depend solely on $\tau_T$ -- 
note that only the synchrotron component, which can be a small
fraction of the total internal luminosity, will be effectively 
scattered by the mirror.

\begin{figure*}[hbt]
\centerline{\epsfig{file=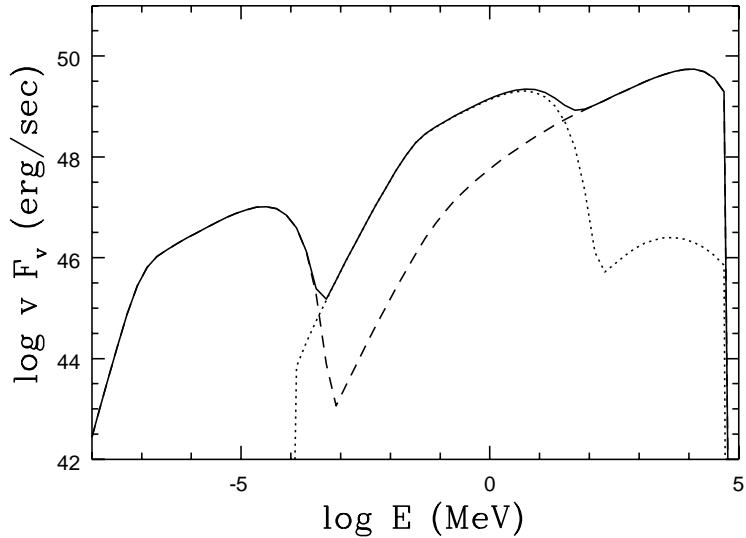,height=12.0cm,width=12cm}}
\caption{Snapshot of the multiwavelength spectrum at the
peak of the direct emission as seen by   
an observer at Earth in the case of the parameters of Fig.~1 and
for $\delta=\Gamma$.
The dashed line corresponds to  the 
directly produced component
 while the dotted line to the doubly
reflected component. The full line is the 
composite spectrum.
\label{growth}} 
\end{figure*}

Figure 3 shows the multiwavelength spectrum at the instant when 
the internal luminosity peaks. The doubly reflected component (dotted line)
peaks in this case at about $5\,(\delta/400)$ MeV, while the directly 
produced synchrotron and inverse Compton components peak respectively 
at $E_{\rm s} \simeq 30(\delta/400)$ eV 
and $E_{\rm ic} \simeq \delta\Gamma~m_ec^2\simeq
80(\delta/400)$ GeV.

\begin{figure*}[hbt]
\centerline{\epsfig{file=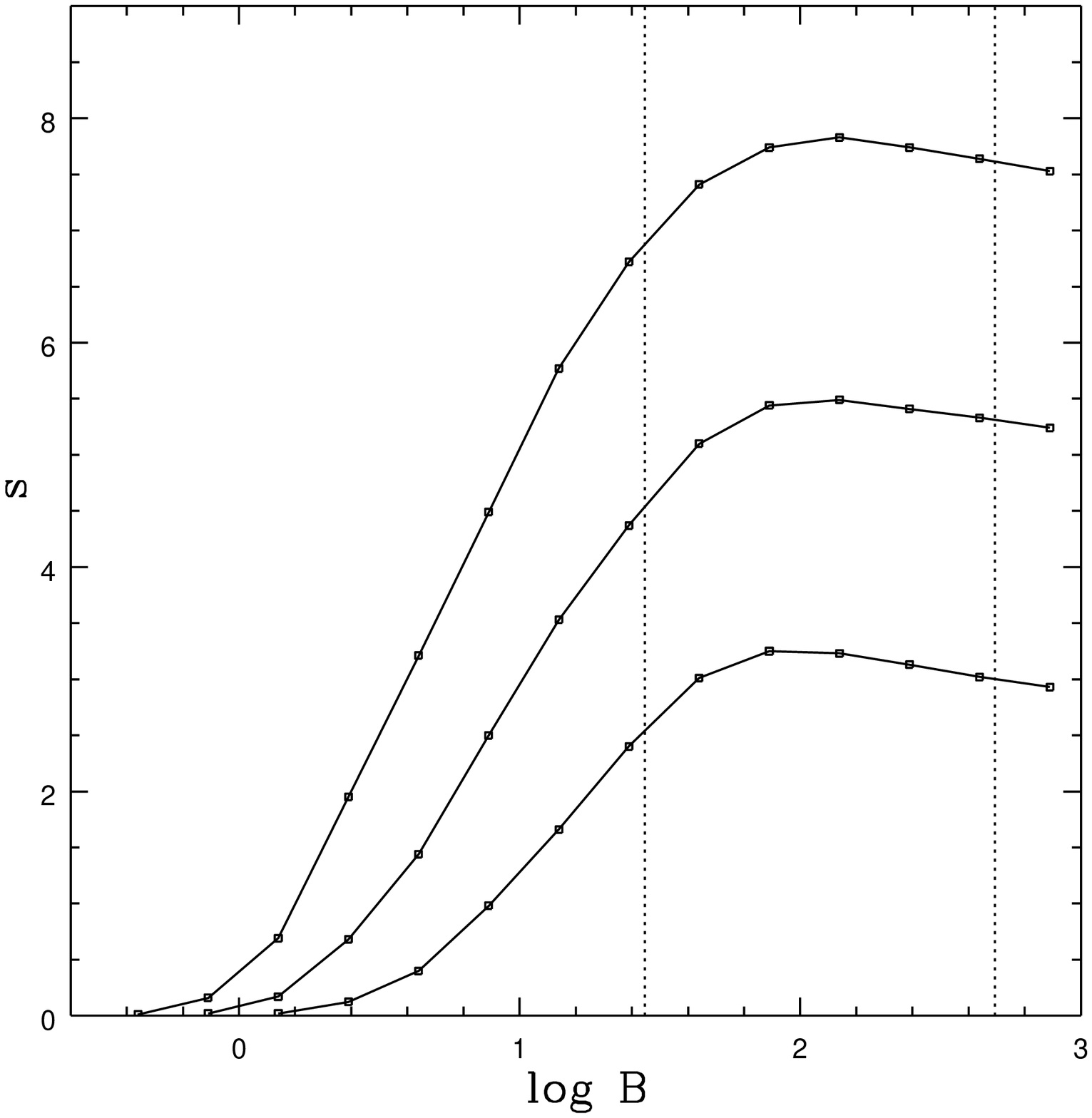,height=9.0cm,width=12cm}}
\caption{Plot of the photon growth index s versus the magnetic field 
strength B (in Gauss) for various values of the proton number density. The
curves from left to right are for 
$n_p\tau_{\rm mir}=10^5,~10^4$ and $10^3~cm^{-3}$.
The rest of the parameters are the same as those of the previous Figures.
The vertical dotted lines represent the two magnetic fields that
correspond to the synchrotron electron
cooling time being equal to one crossing
time for $\gamma=\Gamma$ (left curve) and $\gamma=1.26=\gamma_{min}$ }
\end{figure*}

\section{Modeling the PPS Network -- Including the Effects of Electrons}

In the previous section we outlined the basic ideas of the 
Pair Production/Synchrotron loop in the presence of reflection
(PPSR) emphasizing the fact that the growth of pairs and photons
can take place in the presence of protons alone, i.e. without any 
need for an initial photon or electron populations. In this section we 
use less restrictive, more realistic assumptions by including in 
our calculations an initial population of electrons, an assumption we 
retain for the rest of the paper. Since the plasma must 
be neutral, the least contrived assumption is that the number density 
of initial electrons equals that of the protons (one could also assume the 
additional presence of several $e^+e^-$ pairs per electron but we 
will refrain from doing that  at present). We also note that the energy of these 
electrons as the RBW enters the reflection zone (where all the action
takes place) is essentially a free parameter. The electrons are 
considered to be a part of the flow of the RBW and therefore the 
randomization of their energy is expected to bring them to Lorentz
factors $\gamma_e=\gamma_p=\Gamma$. However, with cooling times 
far shorter than those of protons, they begin cooling immediately upon
their injection into the RBW; by the time the RBW achieves its
terminal Lorentz factor $\Gamma$ the electrons may have attained 
various degrees of cooling and this might occur well before the RBW has 
reached the assumed reflection zone. Because the evolution of our 
model is determined by the number of photons available in the system, 
the presence or not of electrons that can readily produce such photons 
is expected to influence the solution of the PDEs (8) to (10).

In order to avoid introducing an additional set of free parameters 
pertaining to the maximum energy and normalization of the electrons, 
we study the evolution of the system in two extreme cases:

(i) The electrons have totally cooled upon entering the reflection 
zone. Thus we will assume that $n_p=n_e$ while $\gamma_p=\Gamma$ 
and $\gamma_e=1$, and that the photons produced in the cooling
process do not interact with the ``mirror". This case, which we will 
call the `Cooled Electron' 
(CE) case,  uses identical assumptions to the ones of the example 
studied in the previous (no electron) section, with the only difference
being the presence of a minimum value for $\tau_{\rm T}$ related to 
the presence of this initial electron population.

(ii) The electrons have the same Lorentz factor as the protons upon 
entering the reflection zone. 
This would correspond to the case in which the RBW ``sweeps" the 
material between the ``inner engine" and the ``mirror" which now
forms part of the RBW.  The fact that the electrons may have cooled
in the meantime is actually of little importance. What really matters
is the number of electrons swept and the photons that will result
from their cooling. Because of the relativistic speed of the RBW,
these photons are not really lost but they are just a distance $R/\Gamma^2$
ahead of the RBW, which can catch up with them once they scatter in the 
``mirror". In this case we simply let the electrons cool only after 
the RBW has reached close to the ``mirror".
We note that even under this assumption the bulk of the energy of the 
RBW continues to be stored in protons as $U_p/U_e=m_p/m_e$, where $U_p$ 
and $U_e$ are, respectively, the proton and electron total energy content.
We will call this case as the `Energetic Electron' (EE) case.

The cases where the electrons cool in various degrees between
the site the blob first reaches the Lorentz factor $\Gamma$
and the starting point of the reflection zone 
can be considered as intermediate cases.

We have ran the code for various combinations of the values
of the comoving magnetic field strength $B$, the number 
density of the protons and electrons  and of the scattering
depth of the mirror while keeping the
rest of the parameters at their values of the previous section.
Our objective is to study the effects of electron inclusion
on the kinematic and dynamic thresholds as well as the 
energetics and spectral shape of the produced outburst.

As a first step we examine the rate of growth, $s$,
in the CE case. This will help us determine the regime
in the parameter space where the instability operates
efficiently. A first important result is that the
rate of growth remains the same for various combinations of
$\nprot$ and $\taumirror$ as long as their product is constant.
Therefore we examine the behaviour of $s$ for various
combinations of the parameters $B$ and $\nprot\taumirror$; the
results are shown in Figure 4.
We find that the loop cannot operate for sufficiently low 
values of the magnetic field strength $B$, a fact that is in 
quantitative agreement with the discussion given in
Section 2. However we find that the threshold value 
(corresponding to $s = 0$) is not totally independent  
of the product
$\nprot\taumirror$; this can be accounted by the fact that
we use the full single electron synchrotron emissivity 
rather than its delta function approximation
used in KM92 and KGM. We find also that the value
of $\bmin$ is lower from the theoretically found value in 
KGM because the reflected photons are boosted not 
simply by $\Gamma^2$ but by $~4\Gamma^2$ relaxing even more
the threshold requirement. 

 The EE case, because of its very fast development -
the presence of energetic electrons causes the protons 
to lose their energy in one or two crossing times, see Fig. 6 -
and our finite time resolution, does not allow for a similar 
$s~vs.~B$ plot.   As we 
will show later, in this case there are more meaningful ways to check 
the strength of  the instability.

\subsection{Thresholds}

In this section we expand on the results of Fig. 4 by
investigating the full behavior of the photon growth
in time. A study of their growth for various values 
of the comoving magnetic field provides insights on the 
kinematic threshold, while a similar study by varying 
the proton number density provides  insights on the 
effects of the numerically derived dynamic threshold.

{\sl (i) The Kinematic Threshold} 
 
Figures 5 and 6 show the photon lightcurves 
produced after the RBW has entered the 
reflection zone as measured in the comoving frame
for various values of the RBW magnetic field.
Both the proton and the electron number densities
have been kept constant for all runs at
$n_p=n_e=10^4\,{\rm cm}^{-3}$ while $\taumirror=1$.
Fig. 5 shows the case where all electrons have cooled 
completely when the RBW is still in Zone I
(CE case) while the protons have initial energy 
$\gammap=\Gamma=10^{2.6}$. Fig. 6 assumes that
both species are initially at energy 
$\gammap=\gammae=\Gamma=10^{2.6}$ (EE case). 
The B-field ranges from 0.12  to $120$ Gauss with 
increments by a factor of 10 while the rest of the parameters are
the same as those assumed in the example of section 4.

In the CE case (Fig. 5) the photons increase initially 
as $l_\gamma\propto e^{st}$ with $s$ a function of the 
magnetic field strength $B$ (see Fig. 4). At some point 
they reach saturation for the reasons explained
in the previous section. As expected,
for low enough values of $B$
(rightmost curve) the increase is very gradual. 
For even lower $B-$values the photons do not grow at all
in agreement with the kinematic threshold concept discussed
in Section 2. 

The case where the electrons are energetic (EE) before
entering the reflection region (Fig. 6) exhibits some differences
from the CE one. At low $B-$fields we get only the effects 
of the fast cooling electron population with the protons 
remaining practically unaffected. As the electrons cool very 
quickly on their own reflected radiation, the lightcurves peak 
equally fast (i.e. within one or two crossing times) after the
entrance of the RBW inside the reflection zone. As the value 
of $B$ increases the PPSR loop begins to operate and an increasing 
fraction of the proton energy is turned into radiation, as is manifested
by the the higher peak luminosities and the longer decay times of 
the lightcurves.  

These results are summarized in Fig. 7 that depicts the
total photon energy radiated in each run as it 
is measured in the frame of the RBW. As we have already 
mentioned, the energy lost by the protons 
goes into electron-positron pairs and eventually
escapes the system as radiation. Therefore, as the numerical
code discussed in Section 3 conserves energy we expect
\eqb
\int dt L_\gamma(t)=
\Delta U_p + \Delta U_{e,in},
\eqe
where $\Delta U_p$ and $\Delta U_{e,in}$ is the total energy lost by
the protons and initial electrons respectively and 
$L_\gamma$ is the radiated photon luminosity.
In practice, since the code calculates the photon luminosity
at each crossing time $t_i$, the integral in the last relation has
to be replaced by a sum of $\Sigma L_\gamma(t_i) \Delta t_i$.

The dashed and full lines in Fig. 7 represent the CE and EE cases, 
calculated from Fig. 5 and 6 respectively.
At low $B$ there is a marked difference between the two cases:
In the EE case, a substantial fraction of the initial electron
energy is radiated, despite the fact that the system is below
the PPSR threshold. For this reason we get essentially no 
radiation in the CE case that lacks the energetic electrons.
As $B$ increases the PPSR loop begins to operate extracting energy 
from the protons. We note that in the EE case, the presence of 
the initial energetic electron population (and the photons resulting
from their cooling) allows the extraction of energy from the 
protons for lower values of the magnetic field. At even higher 
values of $B$ both cases saturate at the same value of luminosity
as the PPSR loop is capable of cooling the protons completely
irrespective of the presence of initial energetic electrons -- note
however that, according to Figs 5 and 6, the resulting lightcurves 
are different as the time evolution of the system is sensitive 
to the initial conditions (i.e. the presence or not of
relativistic electrons). In this respect, it is of interest to
note that the present model allows for different characteristics 
of the GRB light curves on the basis of the values of its internal
parameters. This is a desirable state of affairs as the light
curve properties themselves could be used, in principle,  to infer or restrict
the values of these parameters. 

\begin{figure*}[hbt]
\centerline{\epsfig{file=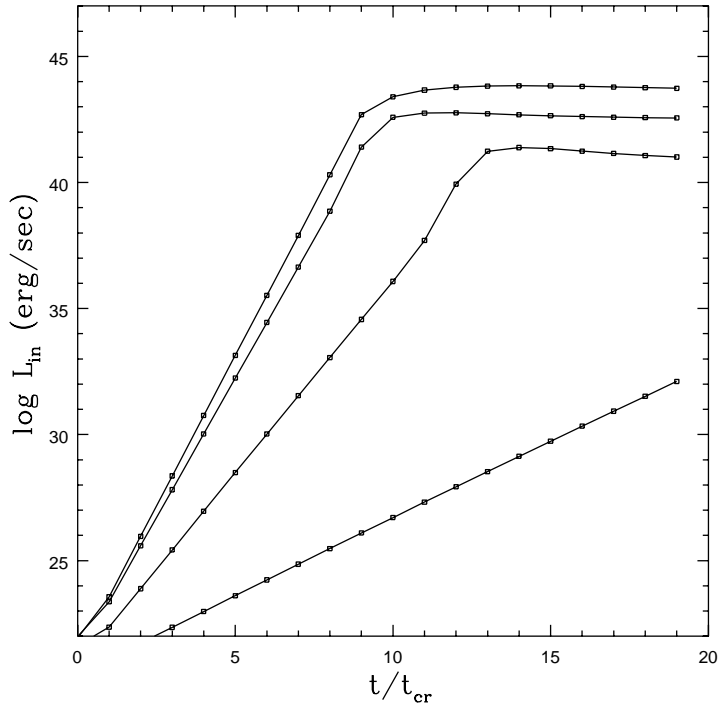,height=10.0cm,width=10cm}} 
\caption{Plot of the photon luminosity evolution as this is
measured in the comoving frame
in the CE case where $n_p=n_e=10^4 \, {\rm cm}^{-3}$
while the value of $B$ has been assumed
4.4, 14, 44 and 140 Gauss (bottom to top). 
The rest of the parameters are as stated in the text. 
Time is measured in blob crossing times and the value $t=0$
has been set at the instant when the RBW enters the
reflection zone.} 
\end{figure*}                                               

\begin{figure*}[hbt]
\centerline{\epsfig{file=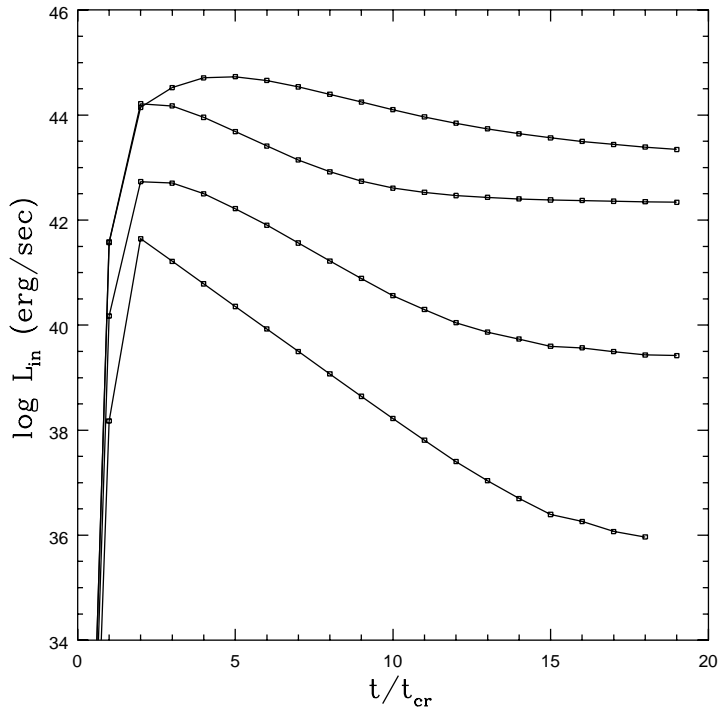,height=10.0cm,width=10cm}} 
\caption{Plot of the photon luminosity evolution
as this is measured in the comoving frame
in the EE case where $n_p=n_e=10^4 \, {\rm cm}^{-3}$ while 
$B$ varies (bottom to top) from 0.12 G to 120 G by increments of
a factor of 10. The rest of the parameters are as stated in the 
text. Time is measured in blob crossing times and the value $t=0$
has been set at the instant when the RBW enters the
reflection zone.}
\end{figure*}                                               

\begin{figure*}[hbt]
\centerline{\epsfig{file=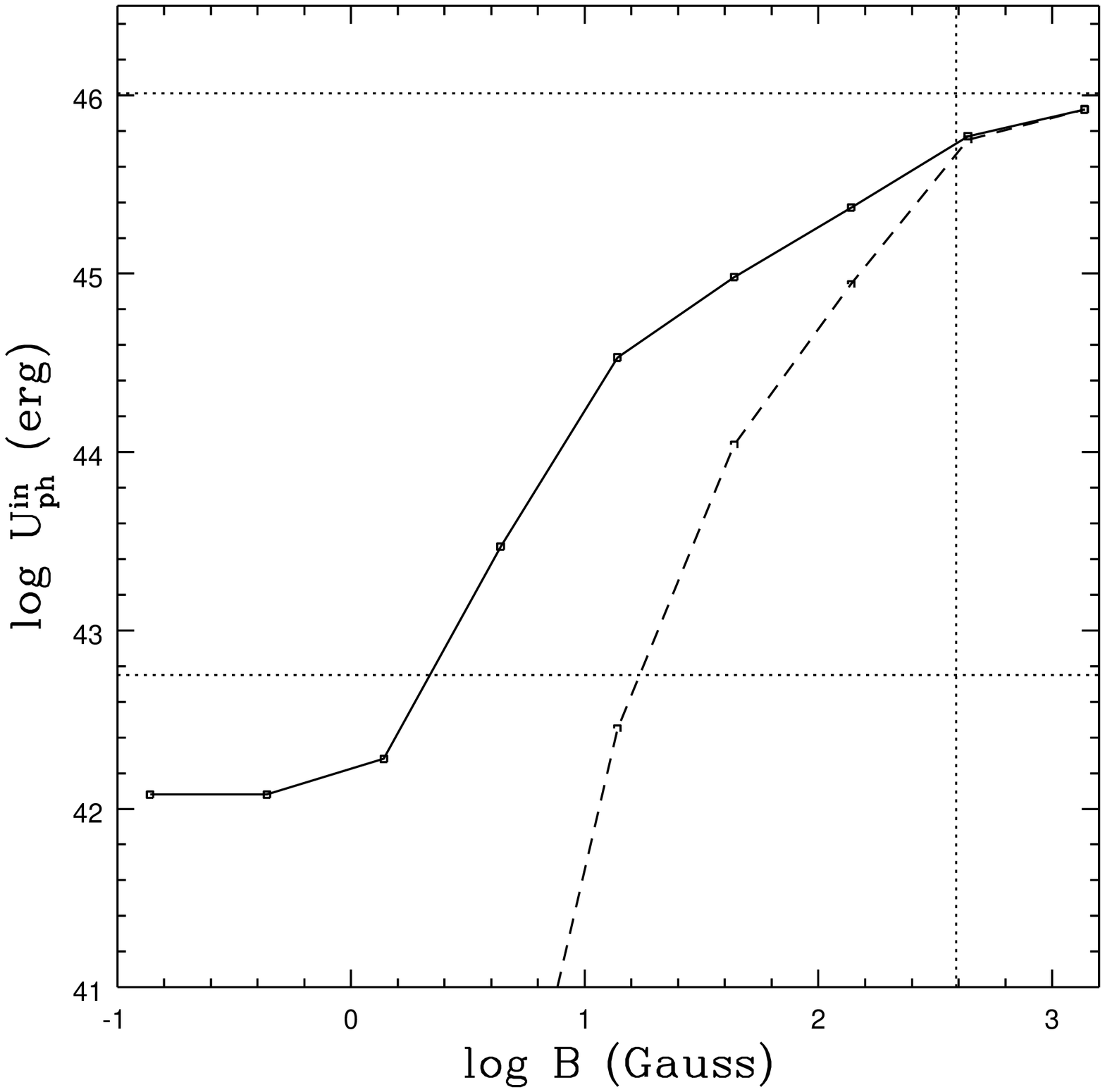,height=10.0cm,width=10cm}} 
\caption{The total energy radiated by the protons 
in the case where $n_p=n_e=10^4 \, {\rm cm}^{-3}$
as a function 
of the magnetic field strength $B$ - both quantities are 
measured in the comoving frame. The other parameters
are given in the text. Full and dashed lines 
represent the EE and CE cases respectively. The lower
horizontal dotted line is the total energy content stored
in electrons while the upper one is the energy stored
in protons. The vertical dotted line is the equipartition
magnetic field.
}                                                
\end{figure*}                                               

{\sl (ii) The Dynamic  Threshold}

To study the dynamical criterion we performed runs
keeping the strength of the comoving field constant
while varying the proton (and electron) number
density. We show the results in Fig. 8. Here we
show the fraction of energy that the protons lose
during the outburst  as a function of $\nprot$.
We find that as $\nprot$ increases the outburst is able
to extract an increasing fraction of the proton energy content.
However the effects are less dramatic than in the
previous case: changing $\nprot$ has an immediate effect
on the efficiency of the PPSR loop but not on its existence.
The loop operates in both the CE and EE cases with 
the EE case again being more efficient in extracting 
energy off the protons. Eventually the two cases converge 
at sufficiently large values of $\nprot$ and their 
efficiency in transfering energy from the protons 
approaches 100 \%. It is worth mentioning at this point
that all the above runs exhibit very similar synchrotron 
spectra, i.e. they all peak at the same energy, as the 
magnetic field has been assumed constant.

\begin{figure*}[hbt]
\centerline{\epsfig{file=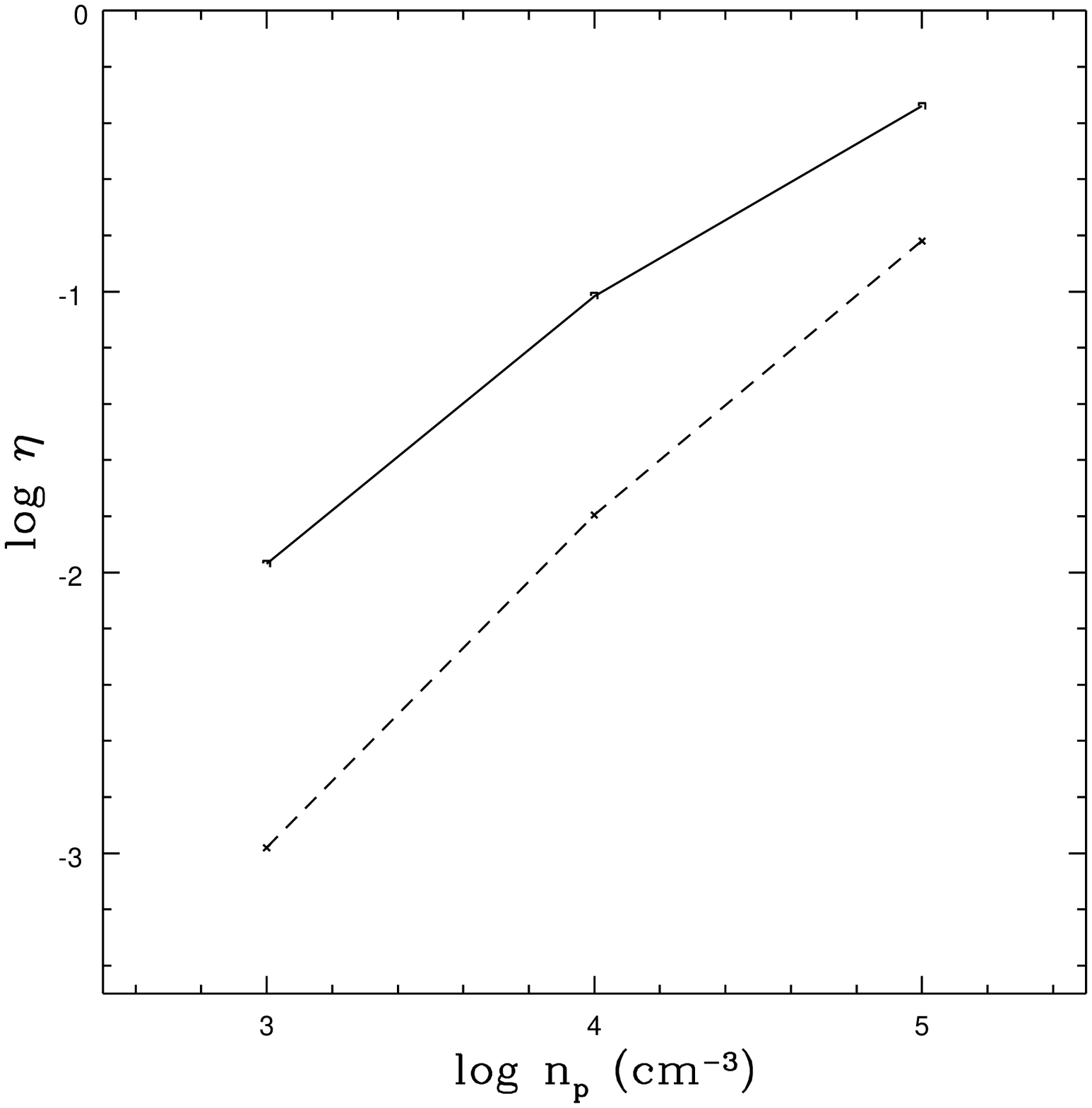,height=10.0cm,width=10cm}} 
\caption{Fraction $\eta$ of the proton energy 
content lost to radiation 
in an outburst as a function of the initial proton
number density. The magnetic field has been assumed to be
 B=120 G. The full line curve corresponds to  the EE case
while the dashed line curve to the CE case. For the other parameters
see text. }                                                
\end{figure*}                                               

\subsection{The Observed Spectra: Direct and
Reflected Components}

We proceed by studying the characteristics of the produced
spectrum, both in the RBW frame as well as observed on Earth. 

We find that in all cases the multiwavelength spectra produced 
on the RBW frame consist of a synchrotron component at low energies 
and an inverse Compton component at high energies. The spectra as 
observed on Earth have a third component  that is produced from the 
double reflection of the aforementioned synchrotron component first 
on the mirror and then on the cooled pairs of the RBW (see Fig.3). 
Next we discuss the characteristics of each component separately. 

{\sl (i) The Synchrotron Component}

Once a value for $\Gamma$ has been assumed, the only important 
parameter for determining the peak of the directly produced synchrotron
spectra, at least during the growth phase, is the value of the 
comoving magnetic field $B$. The peak of the synchrotron component as 
observed in the RBW is given by $\epsilon^{RBW}_{_{S},p}= b\Gamma^2$
and thus  the same quantity as observed at Earth is 
$\epsilon^{obs}_{_{S},p}= \delta b\Gamma^2$ 
(all energies are expressed in units of $m_ec^2$; in the above expressions
and those of the next two subsections
the superscript refers to the frame at which each energy is observed, while
the subscript to the specific process considered with the $p$ after 
the comma referring to the energy of peak emission of the specific process).
One should note here the difference of our model from those more 
common
in the literature for which the electron maximum energy is estimated from
equipartition with protons arguments, yielding much higher energies for the
electrons and synchrotron emission.

{\sl (ii) The Inverse Compton Component}

The peak of the inverse Compton component is largely independent 
of the magnetic field strength $B$, as a substantial fraction of 
the relativistic electron collisions with the 
reflected photons (the dominant contributors to their losses) occur in 
the Klein-Nishina regime. Thus we expect the scattered photons 
to take a substantial fraction of the electron energy (Blumenthal 
\& Gould 1971), leading to $\epsilon^{^{RBW}}_{_{ICS},p}= \eta\Gamma$
and $\epsilon^{obs}_{_{ICS},p}=\eta\delta\Gamma$
with $0.1<\eta<1$.

{\sl (iii) The Doubly Reflected Component}

This component peaks around
$\epsilon^{obs}_{_{R}, p} \simeq \delta\Gamma^2
\epsilon^{^{RBW}}_{_{S},p}$
where the factor $\delta$ comes from the Doppler
blueshift and the factor $\Gamma^2$ comes from the
reflection on the mirror. Using the value of
$\epsilon^{^{RBW}}_{_{S},p}$ given above, we arrive at
$\epsilon^{obs}_{_{R},p}=\delta b\Gamma^4 $
which is basically the relation derived in KGM
(for $\delta=\Gamma$).
As a matter of fact, since  higher values
of $B$ produce higher synchrotron peaks
we expect that Klein-Nishina corrections will
modify the peak of the reflected spectra, therefore
the analytic relation given above
can be used as an upper limit.

\begin{figure*}[hbt]
\centerline{\epsfig{file=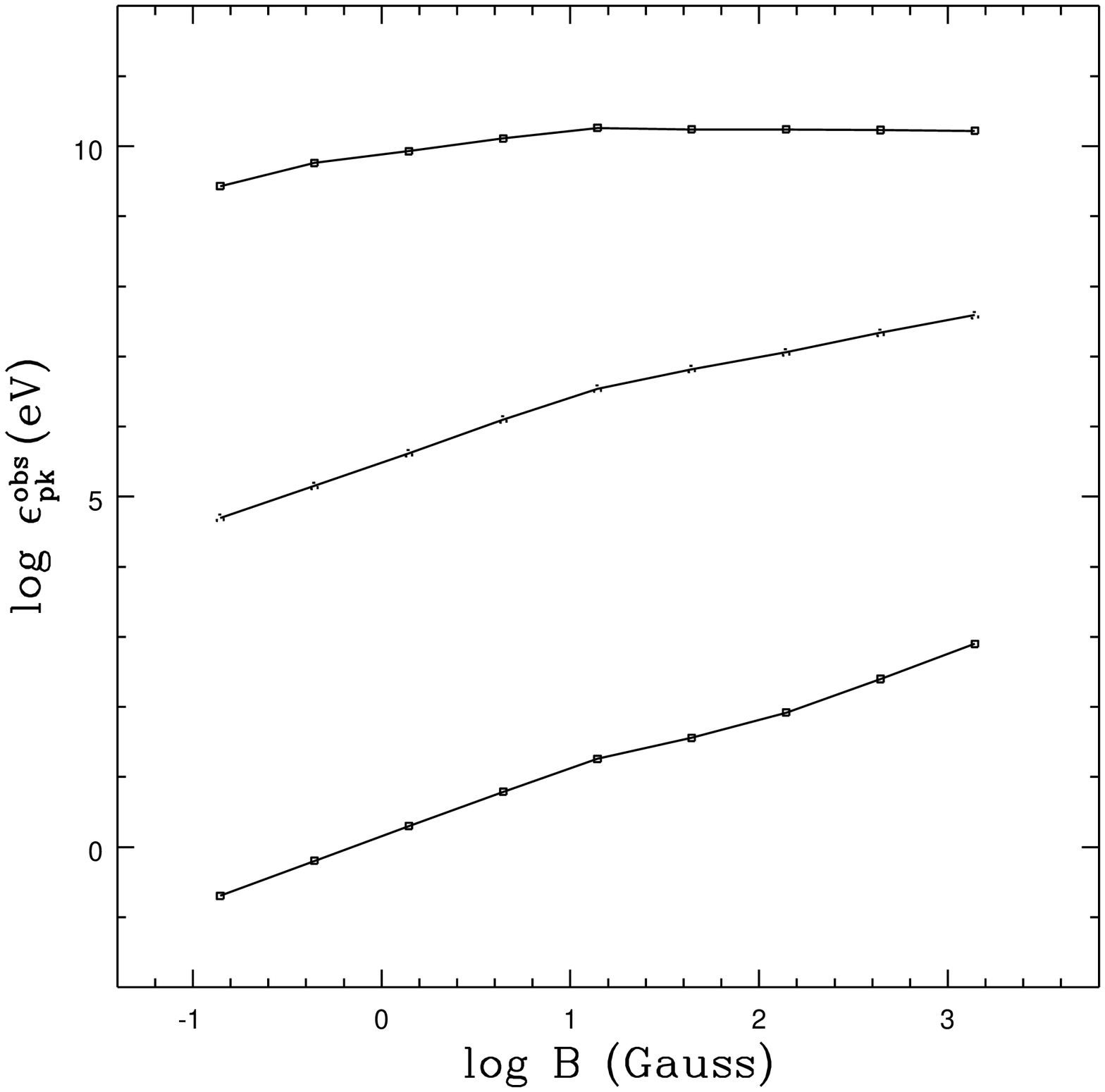,height=10.0cm,width=10cm}} 
\caption{Plot of the peak
of the observed synchrotron (lower curve),
doubly reflected (middle) and inverse Compton (top) component
versus
the comoving magnetic field strength. For the parameters used see text }
\end{figure*}                                               

\begin{figure*}[hbt]
\centerline{\epsfig{file=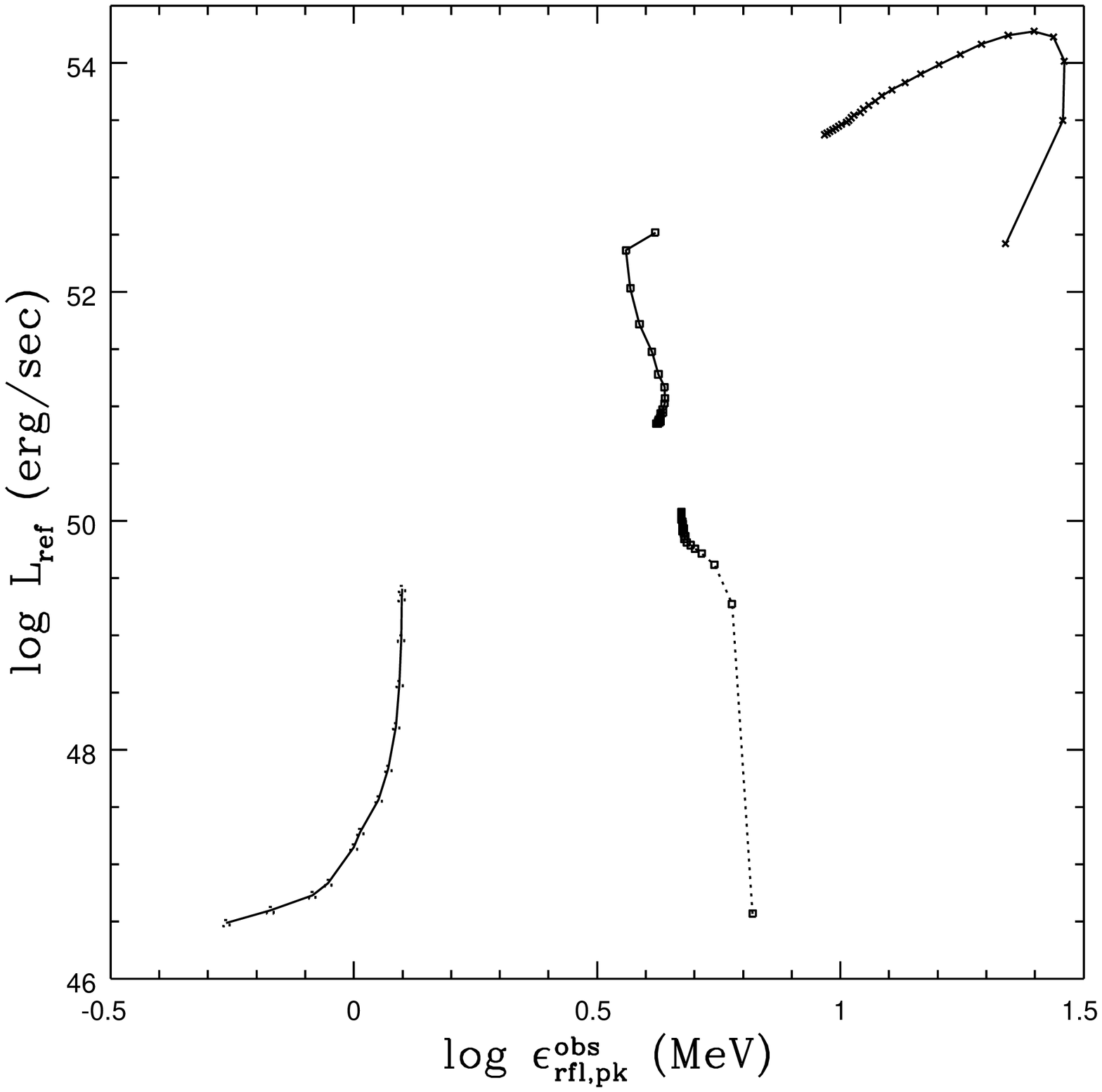,height=10.0cm,width=10cm}} 
\caption{Plot of the evolution of the peak
of the doubly reflected  component
versus the observed luminosity in that component in the EE case and
for various comoving magnetic fields strengths. 
Left curve is for B=4.4 G, middle for B=44 G and right
for B=440 G. For comparison the curve in the CE case for
B=44 G has also been plotted (dotted line). Ticks represent
the crossing times.}
\end{figure*}                                               

Fig. 9 summarizes these results. It shows the
peak energy of the three components as observed on Earth
in the case where $\delta=\Gamma$ versus the 
comoving magnetic 
field strength. We notice that the synchrotron
peak increases linearly with $B$, while the
inverse Compton peak is largely independent 
of it and roughly equal to $\delta \, \Gamma \, m_ec^2$ i.e. 
maximum electron energy as preceived in the lab frame. 
The doubly reflected peak increases with $B$ (as the PPSR loop runs 
increasingly above its kinematic threshold) but the relation is 
slower than linear. Note that for combinations of $\Gamma$
(here taken to be equal to 400) and B that are 
close to threshold (Eq. 3), this peak lies between the 100 keV
- 1 MeV range, and as we emphasized in section 2 this
is independent of the particular choice of $\Gamma$ and B.

However we note that in reality the picture is much more complex.
There is continuous evolution of the spectra due to the time-dependent 
cooling of electrons and protons and we find that  as a result the
frequencies of the peak emission of these components depend on time. 
Fig. 10 depicts the peak of the doubly
reflected component as a function of the luminosity at
each crossing time during the growth phase for various
values of the comoving magnetic field. It is clear that we cannot 
attribute a unique value of the peak energy to 
a certain assumed value of the magnetic field. However,
in order to simpilfy the picture, we can assign as 
a characteristic energy of the peak the energy during
the growth phase and this is what is shown in Fig. 9.
 
In addition to the energy  of peak emission of the three 
components of the spectrum,
we are also interested in their inferred total energy content, 
assuming their emission to be isotropic, a quantity found to 
exhibit several rather tight correlations in GRB statistics
(e.g. Amati et al. 2002). In Section 5.1
we showed the dependence of the total radiated energy 
(as measured in the comoving frame)
on the two important parameters of our problem, namely
the comoving magnetic field strength and the 
number density of the particles. We find that during the 
peak of the burst the relation $\ell_b/ \lphotref \ll 1$ is true
and therefore the bulk of the radiated electron energy will 
be emitted by the 
inverse Compton component, with only a small fraction of it 
radiated by synchrotron. This can be seen also from the test 
case of the previous section (Fig. 3).

However, the luminosity of the doubly reflected 
component has a different time dependence from the luminosity
of the directly observed components shown in Fig. 6, as it 
depends (i) on the synchrotron luminosity of the direct component
and (ii) on the number of cool pairs that have been 
accumulated during the outburst in the RBW, i.e.
on the front's ability to scatter towards our direction
at least a part of the reflected by the mirror radiation.

Fig. 11 depicts the total isotropic energy content of 
each of the burst components, i.e. the synchrotron, the IC, 
and the doubly reflected synchrotron as inferred by an observer  
on Earth as a function of the 
comoving magnetic field strength for the fiducial values
of the rest parameters of our model. The solid line is the 
total directly  radiated energy, which for all practical 
purposes can be considered equal to the energy radiated 
in the inverse Compton component; the dashed line is
the energy radiated by the synchrotron component, while the 
dotted line is the inferred energy associated with the 
doubly reflected component at energy $\sim 1$ MeV and it is 
by and large the component that defines a GRB as such.
The fast increase of this latter component with B 
is due to the increasing number of pairs produced in the
RBW which, in turn, increase its scattering efficiency.

It is of interest to note that for sufficiently small 
values of $B$ the energy in the reflected component 
becomes comparable (or even less) to that of the synchrotron. As of 
today, there have not been any bursts exhibiting prompt optical 
isotropic energy (or, almost equivalently, flux) comparable to that of the 
$\gamma-$rays. In the context of the present model this constraint
sets a lower limit on the value of the comoving field $B$. However, 
besides its dependence on $B$  the ratio of these two components
depends also on the value of $n_p$. While an increase in $n_p$ leads 
to an overall higher photon production rate, it also leads to an increase in 
$\tthom$ which amplifies only the luminosity of the scattered component.

\begin{figure*}[hbt]
\centerline{\epsfig{file=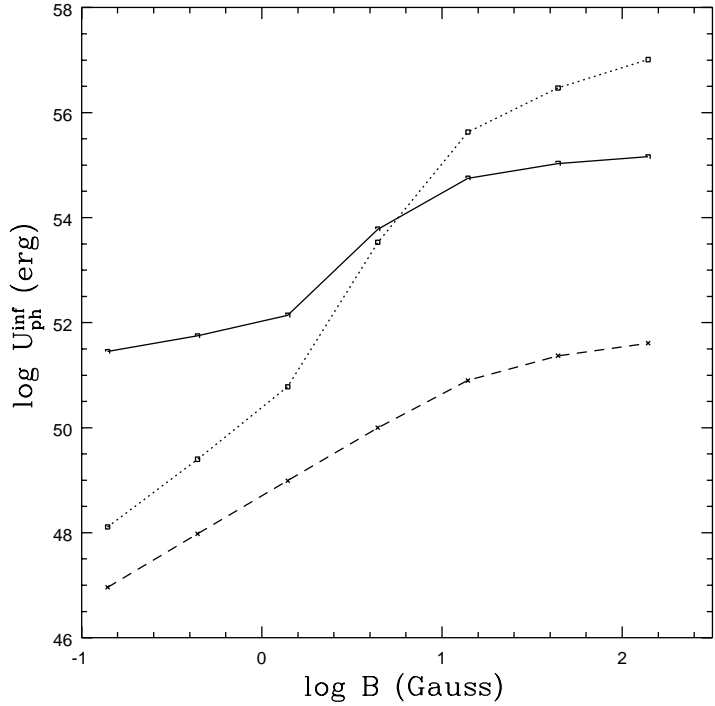,height=10.0cm,width=10cm}}
\caption{Plot of the $4\pi$ inferred energy
content in an outburst
versus
the comovng magnetic field for the EE case. 
The full line curve corresponds
to  the total directly produced energy,
the dashed line curve corresponds to the energy content in the
synchrotron component while the dotted line curve to the
energy content in the doubly reflected component. 
The particle
number density is $n_p=n_e=10^5 \, {\rm cm}^{-3}$. For the rest of the
parameters see text.}                                                
\end{figure*}

Fig. 12 exhibits the isotropic energy content in the doubly
reflected component 
as a function of the
energy of its peak emission (assuming $\delta=\Gamma=400$  and 
the redshift of the burst $z_{_{GRB}} = 0$)
for various values of the proton number density of the RBW
($n_p=10^3,~10^4,~10^5 \, {\rm cm}^{-3}$, bottom to top)
-- and having in mind the possible complications introduced
by the evolution  as these were discussed in connection to Fig. 10.
The figure shows that the energy of this component is a very
strong function of $\epsilon^{obs}_{_{R}, p}$
-- for instance, while $\epsilon^{obs}_{_{R}, p}$ varies
by two orders of magnitude, the energy varies by about nine.
This strong dependence, in the presence of a flux limited
sample results in a peak emission that occurs at approximately 
the same energy, largely independent of the total (inferred) energy 
of the GRB (provided that the observer is located close to the direction
of motion of the RBW).  One should note that this figure is precisely
the so-called Amati relation (Amati et al. 2002) with the axes
interchanged. The relation found by those authors is actually 
much weaker (its equivalent slope is 2 instead of 4.5 
obtained here). With the discovery of the much less luminous 
X-Ray Flashes (XRF), it has been suggested (Lamb et al. 2004) 
that the Amati relation extends to much lower energies  and encompasses 
both classes of transients. The physics behind such a unification
between GRB and XRF is currently uncertain; however, Yamazaki et al. 
(2003) proposed that, similarly  to AGN, the GRB -- XRF unification is  
related to the orientation of the RBW velocity relative to the observer, 
with the XRF being GRB viewed at angles  $\theta > 1/\Gamma$. 
The relation of figure 12 is an intrinsic one and not related to 
the orientation of the observer which is set to $\theta = 1/\Gamma$.

\begin{figure*}[hbt]
\centerline{\epsfig{file=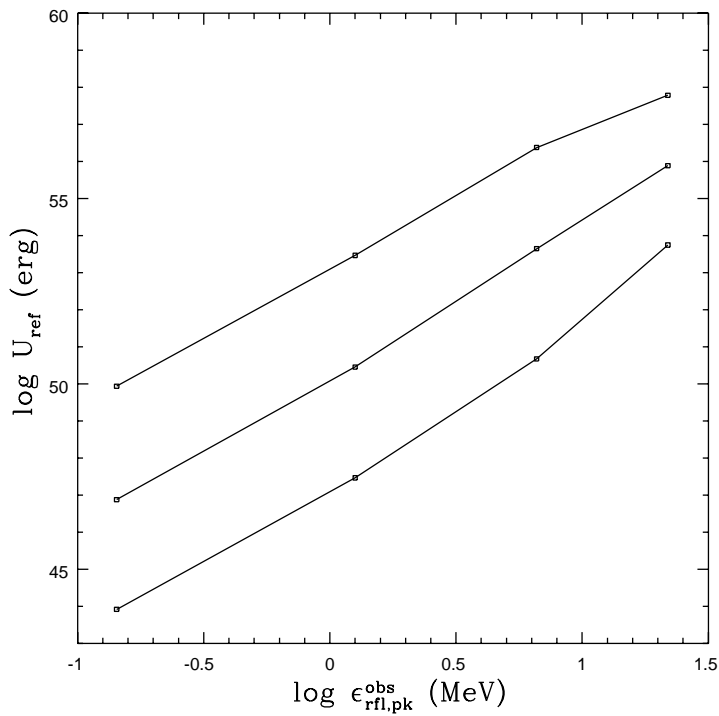,height=10.0cm,width=10cm}}  
\caption{Plot of the  isotropic energy content of the
reflected component as inferred at Earth
versus the peak of
its emission (assuming $\delta=\Gamma=400$)
for various values of the proton number density
($n_p=10^3,~10^4,~10^5\, {\rm cm}^{-3}$, bottom to top).}
\end{figure*}

\section{Summary, Discussion}

In the present paper we have simulated in considerable detail 
the spectral and timing properties of GRB emission within the framework
of the ``Supercritical Pile" model proposed earlier by KGM. 
This model was conceived as a means of resolving two fundamental problems 
of GRB, namely the dissipation and conversion into radiation 
of the energy stored in the protons of the RBW of a GRB and the narrow 
distribution in the energy of the GRB peak emission $E_{\rm p}$.
The ``Supercritical Pile" model resolves both issues with 
a single set of assumptions, making it in this respect unique among 
GRB models. As discussed in KGM, the fundamental process behind
the dissipation of the proton energy is the radiative instability 
discussed in KM92 and MK95, enlarged in scope to include the 
reflection of photons by matter upstream of the advancing RBW, 
a modification that helps reduce  the kinematic and dynamic 
thresholds. Both ingredients are important in producing model spectra 
in general agreement with observations and model parameters consistent
with those thought to prevail in GRB.

The detailed calculations of the present paper show that: (i) The 
presence of reflection reduces both the kinematic and dynamic 
thresholds of the PPSR loop discussed by KM92 in accordance with 
the arguments put forward in KGM. (ii) The process invoked in KGM can 
indeed extract a substantial fraction of the proton energy within 
a few crossing times of the radiating segment of the RBW, for 
judicious choices of the model parameters; for typical values 
of the Lorentz factor 
$\Gamma$, this time scale is in general agreement with GRB 
observations. (iii) The prompt $\nu F_{\nu}$ GRB spectrum comprises 
the following three components (in the observer's frame): (a) A 
broad feature with peak luminosity at energy $E_s \simeq m_e c^2/\Gamma^2$
i.e. at  IR-optical-UV frequencies, the result of synchrotron radiation 
by the pairs produced by the instability. (b) A component with a peak 
at high energies ($E_{\gamma}\simeq \Gamma^2 \, m_e c^2$), 
produced by the inverse Compton scattering of the internally produced
and  reflected radiation 
on the `hot' (i.e. uncooled) pairs of the RBW. (c) A component produced 
by the Compton scattering of the reflected radiation on the 
`cool' pairs that have been accumulated on the RBW. This last 
component peaks in the range of ($0.5-10)\,\delta/\Gamma$
MeV (in the observer's frame), an energy which reflects the 
kinematic threshold of the loop and thus it is largely 
independent of the particular value of $\Gamma$.

The important parameters of our problem are the comoving 
magnetic field strength B, the proton number density
in the RBW frame $n_p$, and the value of the bulk Lorentz factor 
$\Gamma$. Other parameters such as the optical depth of the 
``mirror" and the existence or not of energetic electrons
are less crucial for the formation of the spectra, but of 
importance for the ensuing time evolution. An additional 
parameter which we believe it to be of great importance 
for the observed GRB spectra but which we have not explored at
all as yet is that of the angle between the velocity of the 
`blob' and the observer's line of sight (see e.g. Ioka \& 
Nakamura 2001) which has been proposed as the parameter leading
to the unification between GRB and XRF (Yamazaki et al 2003). 
This parameter proliferation does not pose necessarily a problem
for our (or for that matter any) model, since, as pointed out in 
a recent in-depth analysis of GRB light curve properties, 
at least five parameters are needed to model their diverse 
light curve properties (Norris et al. 2005). 
It is important to stress again here that despite the large
number of available parameters the value of $E_{\rm p}$
is of rather limited range, in agreement with observations.

Our calculations have been done self-consistently by solving 
simultaneously the three space-averaged time-dependent
kinetic equations for protons, electrons and photons,
taking into account all relevant processes between the species
and calculating exactly the reflected photon
component entering at each instant the RBW. 
From all the relevant processes, the ones that
play a key role are Bethe-Heitler pair production,
while synchrotron radiation and inverse Compton scattering 
of electrons-positron pairs are also important  during the
early phases and saturation of the outgrowth
respectively. The rest processes, while included
in the code, are of marginal importance. Thus photon-photon
pair production either does not apply 
(as is the case for collisions
between the directly produced synchrotron
photons and the reflected photons which do not meet 
the threshold requirement) or has a very low optical
depth (as is the case for collisions between
the directly produced photons of the inverse Compton
component and the reflected component ones). 
Therefore, the implied modifications on the photon spectrum
and on the pair  production injection rate are very small
and not able to alter any of our analytical
(and numerically verified) results. Similarly 
we find that electron-positron annihilation 
is negligible as the pairs produced from thr Bethe-Heitler 
process never become optically thick, at least for the
conditions considered herein. Finally, adiabatic
losses have not been taken into account but as the
whole burst episode requires only a few crossing times,
this type of losses  can be neglected.
  
A (perhaps significant) limitation of our calculations
is the assumption of a constant bulk Lorentz factor
$\Gamma$. This might have led to an overestimation of the 
burst luminosity, especially during its decay phase,
as potentially important Compton drag effects and reduction in
the value of $\Gamma$ have not been taken into account. 
However this is not expected to alter our results during 
the growth and peak phases as the protons have not lost yet 
any substantial part of their energy so the evolution under 
a constant $\Gamma$ appears to be a valid hypothesis. 
In this respect, we would like to point out that our model
offers a natural explanation for the termination of the 
prompt phase of the GRB and the onset of the afterglow: this 
comes about when the values of $\Gamma$ and/or 
$B$ drop below the values need to fulfill the kinematic threshold
of our model. At this point no more electron injection takes
place and the GRB enters the afterglow phase characterized 
by the cooling of the available electron population. We 
should point out for the benefit of the reader that in a 
realistic model, the RBW in its propagation sweeps new material 
lying ahead of it, thus increasing its column density. In our 
calculations we have ignored this additional increase in column
density as well as the concommitant decrease in $\Gamma$.

There are several issues that we have not addressed or explored 
in our analysis, mainly because we would like to keep it focused
on the issue of spectral formation and evolution but also 
of finite length. These include:

(1) The origin and dynamics of the ``mirror":
Our discussion to this point assumed that the ``mirror" is 
static relative to the observer, a situation most appropriate, in 
our opinion, to the ``external shock" GRB model. However, given the 
current interest in the alternative ``internal" shock models, 
and also in the possibility that the GRB flux may impart a non-zero
velocity on the mirror matter (Beloborodov 2002), it is easy to generalize
our model to include the effects of the relative motion between 
the RBW and the ``mirror". Such an arrangement will modify the 
kinematic threshold condition from $b\, \Gamma^5 \simeq 2$ to $b 
\,\Gamma^3 \, \Gamma_{\rm rel}^2 \simeq 2$, where $\Gamma_{\rm rel}$ 
is the relative Lorentz factor between the RBW and the ``mirror". 
Arguments very similar to those of Section 2 indicate then, that in 
this case the peak energy of the doubly scattered component will be at 
energy $E_{\rm p} \simeq b \,\Gamma^3 \, \Gamma_{\rm rel}^2 \simeq 2$, 
i.e. it will remain unchanged (it reflects the threhold for pair 
production), as implied by observation.
However, in this case the peak emission of the synchrotron and IC
components will be at energies smaller and larger than $E_{\rm p}$ by 
a factor $\Gamma_{\rm rel}^2$ respectively  rather than $\Gamma^2$. 
Of course, a small value for $\Gamma_{\rm rel}$ will have to be compensated
by a correspondingly larger value for $\Gamma$ in order to fulfill
the kinematic threshold. 
Along the same lines of internal shock models, one might consider  the 
case of a burst of finite duration in which the radiation from its 
earlier ejected segement that has slowed-down, serves as the 
source of photons 
in place of the radiation reflected on the ``mirror". The evolution in 
such a case becomes more complicated because it depends on the details of 
the kinematics of the entire burst front structure. However, assuming 
that the emission from the slowed-down section is again at an energy 
rougly equal to $\epsilon_s \simeq b \Gamma^2$ leads to spectra
very similar to those produced by our present calculations. 

In our work so far we have refrained from discussing the nature
of the ``mirror" invoked in our model.  As far as the kinematic
threshold of our model is concerned, this is independent of the 
``mirror" albedo $\tau_{\rm mir}$. However, the dynamic threshold,
and hence the efficiency of radiating away the proton energy, does
depend on this parameter as indicated in Eq.(4). Assuming the 
same column density for the scattering in the ``mirror" 
and using that the ratio 
of the Thomson to the $p \, \gamma$ reaction cross-section
is about $300$ (true for values of the collision energy about 10 times 
above threshold)
we obtain from Eq. (4) the condition $\tau_{\rm mir} \gsim
4 \, \Gamma^{-2}$ which is
satisfied for $\Gamma \gsim 430$
when the density and radius are 
$n \simeq 10^3~{\rm cm^{-3}}$ and $ R_{16} = 3$ respectively. 
The
condition on the density and radius are consistent with a wind of 
$\dot M \simeq 3 \times 10^{-6}\; {\rm M}_{\odot} \, {\rm yr}^{-1}$ and 
velocity $v \simeq 10^8 \; {\rm cm/s}$. At this point we would like to 
re-iterate that fulfilling only the kinematic (but not the dynamic) threshold 
would still result in a burst, however one which is rather 
inefficient
in converting the proton energy into radiation.
Finally, an intriguing possibility is the increase of the number of 
pairs ahead of the shock by the scattering of the high energy photons
and their conversion into electron-positron pairs, as discussed
by Beloborodov (2002), who indicates that, under certain circumstances, 
it may be possible to obtain a pair depth as high as $\simeq 1$.

(2) Exploring the $B, ~\Gamma, ~n_p$ parameter space for 
correlations with GRB phenomenology: 
Our present models were all run with a given value of 
$\Gamma$. Our results indicate the time evolution to be 
faster for combinations of $B, \Gamma$ well above the 
kinematic threshold; at the same time, however, the value of 
$E_{\rm p}$ is larger for such combinations, in agreement
with the general observation that faster varying bursts are
generally harder than slower varying ones. Lower values of 
$\Gamma$ can be compensated by larger values of $n_p$, which
however lead to different values of $E_{\rm iso}$  or peak
luminosity.

Finally, we should re-iterate that, contrary to most, our model 
produces spectra in general agreement with observation without 
the need to invoke the 
presence of shock accelerated particle populations. The
presence of such populations is not excluded, neither would 
invalidate any of the present results; however, it would lead
to spectra more complicated than those produced in this paper
that, in addition, extend to energies higher than those suggested 
in this work. 
One can argue by simple inspection of the threshold conditions
(which can now be fulfilled for much lower values of $\Gamma$)
that the presence of a non-thermal  population of relativistic 
protons would result in emission of high energy 
radiation long after the prompt GRB emission at $\simeq 1$ MeV
has died out. Apparently, there has been at least one such 
event so far, i.e. GRB 941017, registered in the NaI calorimetric 
detectors of EGRET aboard CGRO (Gonz\'alez et al. 2003). It is 
also of interest to note that the highest photon associated 
with GRB emission came approximately 90 minutes after the
end of the prompt emission  in GRB 940217 (Hurley et al. 1994).
The impending launch of GLAST with 
its superior sensitivity may lead to the discovery of more
similar events, which will test the extension of the particle
distributions in GRB to energies higher than considered in this
note.  

The model presented herein makes several concrete predictions.
Thus we expect that, if the $\sim 1$ MeV GRB are produced in 
the way prescribed by this model,
they should be accompanied by prompt emission in the
IR-optical band (the synchrotron component of the
direct emission), as well as in the GeV-TeV regime (inverse 
Compton component). The precise energies of the above peaks 
depend on $\Gamma$ and the comoving magnetic field, however 
the model predicts that GRB should be strong emitters
at energies $\sim \Gamma^2$ (in $m_ec^2$ units). This is
from the inverse Compton component which, during the
peak of the burst, dominates the synchrotron component
by many orders of magnitude. Thus GRB should be abundantly 
detected by both the GBM and the LAT instruments aboard
GLAST, and possibly by ground based TeV telescopes. 

To date there have been a couple of GRB with prompt optical
emission. The first one was that of GRB 990123 detected 
by ROTSE, with optical luminosity $\simeq 10^{-3}$ of that
of the $\gamma-$ray band detected by BATSE aboard CGRO. 
Based on the fact that its optical light curve did not 
follow the detailed shape of the BATSE one, it has been argued
that emission in this band is due to synchrotron radiation by
the reverse shock of the RBW (Sari \& Piran 1999). Interestingly,
synchrotron emission by this component scales $\propto \Gamma^2$
(see Piran 2004, Eq. (72)) just like the synchrotron emission of 
our model. The second burst with optical prompt emission was 
that of GRB 041219a, detected by {\sl Swift} in  $\gamma-$rays
and by the RAPTOR ground based system (Vestrand et al. 2005) in the 
optical band.
Contrary to the case of GRB 990123, the optical light curve
of  GRB 041219a did follow the details of the time evolution of 
its $\gamma-$ray light curve, in agreement with the tenets of the 
present model. It is also of interest to note, that in this case 
too the optical luminosity was roughly $\simeq 10^{-3}$ of the 
high energy one, while the extrapolation of its optical spectrum
did not match that of the high energy one, suggesting two 
spectrally distinct components. The presence of {\sl Swift} in
orbit guarantees that there will be more GRB with prompt emission 
from IR - optical to X-rays to $\gamma-$ray that will help 
determine the viability or not of our model.

Also of interest
are observations by INTEGRAL as the peak of the doubly reflected
component occurs within the INTEGRAL observing energy band, 
a feature that can allow direct comparison between our model and 
observations. To
conclude we only mention the possibility of neutrino emission
within the present scenario. This would be possible if the 
Lorentz factor of the RBW were sufficiently high to lead to 
pion photo-production. This emission could be present both 
in the ``prompt" GRB phase, produced as suggested above, or 
by an accelerated proton component after the end of the prompt
phase, provided that the proton distribution extends to 
sufficiently large values that fulfill the kinematic threshold 
condition.

Finally, we would like to point attention to possible polarization
signatures of the prompt GRB emission in the component comprising 
$E_{\rm p}$ within our model: Because within our model this component 
is due to Compton scattering (rather that synchrotron emission as in 
most models) it should be highly polarized (up to 100\%) if, as believed, 
it represents emission  viewed at angles $\theta \simeq 1/\Gamma$.
The highly polarized emission of GRB 031206 detected by RHESSI
(Coburn \& Boggs 2003), while not totally conclusive adds one more piece 
of evidence in agreement with the model presented above.

{\sl Acknowledgments: AM would like to acknowledge
partial support from 
the Special Funds for Research of the University of Athens
and from the joint Greek-EU "Pythagoras" research grants.
DK would like to acknowledge support by 
an INTEGRAL GO grant. 
We would like also to thank an anonymous
referee for many constructive comments.}

\end{document}